\documentclass[proceedings]{JHEP3a}
\usepackage{eurosym}
\usepackage{amssymb}
\usepackage{amsfonts}
\usepackage{amsmath,epsfig}
\usepackage{graphicx}

\newbox\mybox

\newcommand\fverb{\setbox\mybox=\hbox\bgroup\verb}
\newcommand\fverbdo{\egroup\medskip\noindent\fbox{\unhbox\mybox}\ }
\newcommand\fverbit{\egroup\item[\fbox{\unhbox\mybox}]}
\conference{Goldstone bosons in different PT-regimes of non-Hermitian scalar QFT}
\abstract{We study the interplay between spontaneously breaking global continuous and discrete antilinear 
symmetries in a newly proposed general class of non-Hermitian quantum field theories containing a mixture of complex and real scalar fields.
We analyse the model for different types of global symmetry preserving and breaking vacua. In addition, the models are symmetric under various types of discrete antilinear 
symmetries composed out of nonstandard simultaneous charge conjugations, time-reversals and parity transformations; CPT. While the global symmetry governs the existence of massless Goldstone bosons, 
the discrete one controls the precise expression of the Goldstone bosons 
in terms of the original fields in the model and its physical regimes. 
We show that even when the CPT-symmetries are broken on the level of the action expanded around different types of vacua, the mass spectra might still be real when the symmetry 
is preserved at the tree approximation and the breaking only occurs at higher order. We discuss the parameter space of some of the models in the proposed class and identify physical regimes in which massless Goldstone bosons 
emerge when the vacuum spontaneously breaks the global symmetry or equivalently
when the corresponding Noether currents are conserved. The physical regions 
are bounded by exceptional points in different ways. There exist special points in parameter space for which massless bosons may occur already before breaking the global symmetry. However, when the 
global symmetry is broken at these points they can no longer be distinguished from genuine Goldstone bosons.}

\title{Goldstone bosons in different PT-regimes of non-Hermitian scalar
quantum field theories}
\author{Andreas Fring and Takanobu Taira \\
Department of Mathematics, City University London,\\
Northampton Square, London EC1V 0HB, UK \\
E-mail: a.fring@city.ac.uk, takanobu.taira@city.ac.uk}

\input{tcilatex}
\begin{document}

\section{Introduction}

It is quite well understood how to extend the conventional framework of
Hermitian classical and quantum mechanics \cite{Bender:1998ke,Alirev,PTbook}
to allow for the inclusion of non-Hermitian systems. When the latter systems
admit an antilinear symmetry \cite{EW}, such as for instance being invariant
under a simultaneous reflection in time and space, referred to as $\mathcal{%
PT}$-symmetry, this can be achieved in a self-consistent manner. In these
circumstances one encounters three types of regimes with qualitatively
different behaviour, a $\mathcal{PT}$-symmetric phase, a spontaneously
broken $\mathcal{PT}$-symmetric phase and a completely $\mathcal{PT}$%
-symmetry broken phase. Based on the formal analogy between the Schr\"{o}%
dinger equation and the propagation of light in the paraxial approximation
described by the Helmholtz equation many of the findings obtained in the
quantum mechanical description have been confirmed experimentally and
further developed in classical optical settings with the refractive index
playing the role of a complex potential \cite%
{Muss,MatMakris,Guo,ruter2010obs,el2018non}.

When implementing and extending these idea and principles to quantum field
theories there is less consensus, and for some aspects alternative
resolutions have been proposed. Naturally, as a direct extension of the well
studied purely complex cubic potential in quantum mechanics the scalar field
theory with imaginary cubic self-interaction term $i\phi ^{3}$ has been
investigated at first \cite{benderphi31,shalabyphi31} and also the more
generally deformed harmonic oscillator has been generalised to a field
theoretical interaction term $\phi ^{2}(i\phi )^{\varepsilon }$ more
recently \cite{bender2018p}. Non-Hermitian versions with a field theoretic
Yukawa interaction \cite%
{alexandre2015non,rochev2015hermitian,korchin2016Yuk,laureyukawa} have been
investigated in regard to Higgs boson decay. Besides bosonic theories also
generalizations to non-Hermitian fermion theories such as a free fermion
theory with a $\gamma _{5}$-mass term or the massive Thirring model have
been proposed \cite{bender2005dual}. $\mathcal{PT}$-symmetric versions of
quantum electrodynamics have been studied \cite%
{bender1999nonunitary,milton2013pt} as well.

Here we will focus on a feature that is very central to standard Hermitian
quantum field theory, the Goldstone theorem, and investigate further how it
extends to non-Hermitian theories. We recall that in the Hermitian case the
theorem states that the number of massless Goldstone bosons in a quantum
field theory is equal to the dimension of the coset $G/H$, with $G$ denoting
a global continuous symmetry group of the action and $H$ the symmetry group
that is left when the theory is expanded around a specific vacuum \cite%
{nambu1961dynamical,goldstone1961field}. The question of extension was
recently addressed by Alexandre, Ellis, Millington and Seynaeve \cite%
{alexandre2018spontaneous} and separately by Mannheim \cite%
{mannheim2018goldstone}. Interestingly, both groups found that the theorem
appears to hold for non-Hermitian theories as well, but they proposed two
alternative variants for it to be implemented.\textit{\ }In addition,
Mannheim suggests that the non-Hermitian theory possess the new feature of
an unobervable Goldstone boson at a special point. Here we find that the
Goldstone bosons takes on different forms depending on whether the theory is
in the $\mathcal{CPT}$-symmetric regime, at standard exceptional point or
what we refer to as the \emph{zero-exceptional point}. We distinguish here
between a standard exceptional point, corresponding to two nonzero
eigenvalues coalescing, and a zero-exceptional point defined as the point
when a zero eigenvalue coalesces with a nonzero eigenvalue.

The problem that both groups have tried to overcome at first is the feature
that the equations of motion obtained from functionally varying the action
with respect to the scalar fields on one hand and on the other separately
with respect to its complex conjugate field are not compatible. This is a
well known conundrum for non-Hermitian quantum field theories and has for
instance been pointed out previously and elaborated on in \cite%
{alexandrefoldy,alexandre2017symmetries} for a non-Hermitian fermionic
theory. Hence, without any modifications the proposed non-Hermitian quantum
field theories appear to be inconsistent. To resolve this problem the
authors of \cite{alexandre2018spontaneous} proposed to use a non-standard
variational principle by keeping some non-vanishing surface terms. In
contrast, Mannheim \cite{mannheim2018goldstone} utilizes the fact that the
action of a theory can be altered without changing the content of the theory
as long as the equal time commutation relations are preserved, see e.g. \cite%
{bender2005dual}. Utilizing that principle he investigates a model based on
a similarity transformed action of the previous one in which the entire set
of equations of motion have consistent properties. Remarkably, it was found
for both versions that the theory expanded around the global $U(1)$-symmetry
breaking vacuum contains a massless Goldstone boson. Moreover, while in the
approach that only validates half of the standard set of equations of motion
non-standard currents are conserved and Noether's theorem seems to be
evaded, the approach proposed in \cite{mannheim2018goldstone} is based on
the standard variational principle leading to standard Noether currents.

Here we largely adopt the latter approach and analyse theories expanded
about different types of vacua, global symmetry breaking and also preserving
ones, for a class of models containing a mixture of several types of complex
scalar of fields and also real self-conjugate fields. In particular, we
identify the physical regions in parameter space by demanding the masses to
be non-negative real-valued in order to be physically meaningful. This has
not been considered previously, but is in fact quite essential as
potentially the theory might be entirely unphysical. As is turns out, in
many scenarios we are able to identify some physical regimes that are,
however, quite isolated in parameter space. We find some vacua that break
the $\mathcal{CPT}$-symmetries on the level of the action, but still possess
physically meaningful mass spectra, as the symmetry breaking occurs at
higher order couplings than at the tree approximation. Moreover, we derive
the explicit forms of the Goldstone boson in all three $\mathcal{PT}$%
-regimes, the symmetric and spontaneously broken phases, as well as at the
exceptional point.

Our manuscript is organised as follows: In section 2 we introduce a general
model with $n$ scalar field that might be genuinely complex but in some
versions also contain real self-conjugate fields. In section 3 and 4 we
investigate two specific examples of this general class of models in more
detail and identify the physical regions in which Goldstone bosons may or
may not occur. We investigate different types of vacua that may break the
global $U(1)$-symmetry and also several variants of discrete $\mathcal{CPT}$%
-symmetries that might be broken separately. Starting\ from a complex
squared mass matrix we construct the $\mathcal{P}$-operator that together
with $\mathcal{T}$-operator can be used to identify the real eigenvalue
regime and show how these operators, that can be thought off as quantum
mechanical analogues, are related to the quantum field theoretical $\mathcal{%
CPT}$-operator. We identify the explicit form of the Goldstone boson in
terms of the original fields in the action in different $\mathcal{PT}$%
-regimes. In section 5 we investigate how the interaction term may be
generalised so that the action still respects a discrete $\mathcal{CPT}$%
-symmetry and a continuous global $U(1)$-symmetry. We state our conclusions
and present an outlook in section 6.

\section{A non-Hermitian model with $n$ complex scalar fields}

We consider here generalizations of the model originally proposed in \cite%
{alexandre2018spontaneous} and further studied in \cite%
{mannheim2018goldstone}. To be a suitable candidate for the investigation of
the non-Hermitian version of Goldstone's theorem the model should be not
invariant under complex conjugation, possess a discrete $\mathcal{CPT}$%
-transformation symmetry and crucially be invariant under a global
continuous symmetry. The actions $\mathcal{I}_{n}=\int d^{4}x\mathcal{L}_{n}$
involving the Lagrangian densities functional of the general form 
\begin{equation}
\mathcal{L}_{n}=\sum\limits_{i=1}^{n}\left( \partial _{\mu }\phi
_{i}\partial ^{\mu }\phi _{i}^{\ast }+c_{i}m_{i}^{2}\phi _{i}\phi _{i}^{\ast
}\right) +\sum\limits_{i=1}^{n-1}\kappa _{i}\mu _{i}^{2}\left( \phi
_{i}^{\ast }\phi _{i+1}-\phi _{i+1}^{\ast }\phi _{i}\right)
-\sum\limits_{i=1}^{n}\frac{g_{i}}{4}(\phi _{i}\phi _{i}^{\ast })^{2}
\label{actionan}
\end{equation}%
possess all of these three properties. The parameter space is spanned by the
real parameters $m_{i},g_{i},\mu _{i}\in \mathbb{R}$ and $c_{i},\kappa
_{i}=\pm 1$. The latter constants might be absorbed into the mass and the
couplings $\mu _{i}$ when allowing them to be purely imaginary or real.
However, we keep these constants separately since their values distinguish
between different types of qualitative behaviour as we shall see below. When
fixing those constants to specific values the action $\mathcal{I}_{2}$
reduces to the model discussed in \cite%
{alexandre2018spontaneous,mannheim2018goldstone}. In order to keep matters
as simple as possible in our detailed analysis, we will set here $g_{i}=0$
for $i\neq 1$, but in section 5 we argue that the interaction term may be
chosen in a more complicated way with all three properties still preserved.

Functionally varying the action $\mathcal{I}_{n}$ separately with respect to 
$\phi _{i}$ and $\phi _{i}^{\ast }$ gives rise to the two sets of equations
of motion%
\begin{equation}
\frac{\delta \mathcal{I}_{n}}{\delta \phi _{i}}=\frac{\partial \mathcal{L}%
_{n}}{\partial \phi _{i}}-\partial _{\mu }\left[ \frac{\partial \mathcal{L}%
_{n}}{\partial \left( \partial _{\mu }\phi _{i}\right) }\right] =0,\qquad 
\frac{\delta \mathcal{I}_{n}}{\delta \phi _{i}^{\ast }}=\frac{\partial 
\mathcal{L}_{n}}{\partial \phi _{i}^{\ast }}-\partial _{\mu }\left[ \frac{%
\partial \mathcal{L}_{n}}{\partial \left( \partial _{\mu }\phi _{i}^{\ast
}\right) }\right] =0.  \label{EoM}
\end{equation}%
We comment below on the compatibility of these equations. Evidently, the
action $\mathcal{I}_{n}$ is not Hermitian when $\phi _{i}^{\ast }\neq \phi
_{i}$ for some $i$. However, it is invariant under two types of $\mathcal{CPT%
}$-transformations%
\begin{equation}
\mathcal{CPT}_{1}:\phi _{i}(x_{\mu })\rightarrow (-1)^{i+1}\phi _{i}^{\ast
}(-x_{\mu }),\quad \mathcal{CPT}_{2}:\phi _{i}(x_{\mu })\rightarrow
(-1)^{i}\phi _{i}^{\ast }(-x_{\mu }),~~~~i=1,\ldots ,n.  \label{CPT12}
\end{equation}%
As pointed out in \cite{mannheim2018antilinearity} these types of symmetries
are not the standard $\mathcal{CPT}$ transformations as some of the fields
are not simply conjugated and $\mathcal{P}$ does not simply act on the
argument of the fields, but also acquire an additional minus sign as a
factor under the transformation. Such type of symmetries were studied in the
quantum field theory context in more detail in \cite%
{mannheim2018antilinearity} and as argued therein make the non-Hermitian
versions good candidates for meaningful and self-consistent quantum field
theories, in analogy to their quantum mechanical versions, despite being
non-Hermitian.

In addition, the action related to (\ref{actionan}) is left invariant under
the continuous global $U(1)$-symmetry%
\begin{equation}
\phi _{i}\rightarrow e^{i\alpha }\phi _{i},\quad ~~\phi _{i}^{\ast
}\rightarrow e^{-i\alpha }\phi _{i}^{\ast },~~~~~\quad i=1,\ldots ,n\text{, }%
\alpha \in \mathbb{R},  \label{U11}
\end{equation}%
when none of the fields in the theory is real, that is when $\phi _{i}^{\ast
}\neq \phi _{i}$ for all $i$. Applying Noether's theorem and using the
standard variational principle for this symmetry one obtains 
\begin{equation}
\delta \mathcal{L}_{n}=\partial _{\mu }\left[ \sum\limits_{i=1}^{n}\frac{%
\partial \mathcal{L}_{n}}{\partial \left( \partial _{\mu }\phi _{i}\right) }%
\delta \phi _{i}+\frac{\partial \mathcal{L}_{n}}{\partial \left( \partial
_{\mu }\phi _{i}^{\ast }\right) }\delta \phi _{i}^{\ast }\right]
+\sum\limits_{i=1}^{n}\left[ \frac{\delta \mathcal{I}_{n}}{\delta \phi _{i}}%
\delta \phi _{i}+\frac{\delta \mathcal{I}_{n}}{\delta \phi _{i}^{\ast }}%
\delta \phi _{i}^{\ast }\right] .
\end{equation}%
Thus provided the equations of motion in (\ref{EoM}) hold, and $\delta 
\mathcal{L}_{n}=0$ when using the global $U(1)$-symmetry in the variation
with $\delta \phi _{j}=i\alpha \phi _{j}$ and $\delta \phi _{j}^{\ast
}=-i\alpha \phi _{j}^{\ast }$, we derive the Noether current associated to
this symmetry as 
\begin{equation}
j_{\mu }=i\alpha \sum\nolimits_{i}\left( \phi _{i}\partial _{\mu }\phi
_{i}^{\ast }-\phi _{i}^{\ast }\partial _{\mu }\phi _{i}\right) .  \label{JN}
\end{equation}%
Below we discuss in more detail under which circumstances this current is
conserved. We will argue that Noether's theorem holds in its standard form
and is not evaded as concluded by some authors. Next we are mainly
interested in the study of mass spectra resulting by expanding the
potentials around different vacua as this probes the Goldstone theorem.

\section{Discrete antilinear and continuous global symmetry}

We now discuss the model $\mathcal{I}_{3}$ in more detail with all fields
being genuinely complex scalar fields, i.e. $\phi _{i}\neq \phi _{i}^{\ast }$%
, $i=1,2,3$. Then the action for (\ref{actionan}) takes on the form 
\begin{equation}
\mathcal{I}_{3}\mathcal{(}\phi _{i},\phi _{i}^{\ast },\partial _{\mu }\phi
_{i},\partial _{\mu }\phi _{i}^{\ast }\mathcal{)}=\int d^{4}x\mathcal{L}_{3},
\label{I1}
\end{equation}%
with Lagrangian density functional%
\begin{equation}
\mathcal{L}_{3}\mathcal{=}\sum\limits_{i=1}^{3}\partial _{\mu }\phi
_{i}\partial ^{\mu }\phi _{i}^{\ast }-V_{3},
\end{equation}%
and potential%
\begin{equation}
V_{3}\mathcal{=}-\sum\limits_{i=1}^{3}c_{i}m_{i}^{2}\phi _{i}\phi _{i}^{\ast
}+c_{\mu }\mu ^{2}\left( \phi _{1}^{\ast }\phi _{2}-\phi _{2}^{\ast }\phi
_{1}\right) +c_{\nu }\nu ^{2}\left( \phi _{2}\phi _{3}^{\ast }-\phi _{3}\phi
_{2}^{\ast }\right) +\frac{g}{4}(\phi _{1}\phi _{1}^{\ast })^{2}.
\end{equation}%
Compared to (\ref{actionan}) we have simplified here the interaction term by
taking $g_{1}=g$ and $g_{1}=g_{2}=0$. The model contains the real parameters 
$m_{i},\mu ,\nu ,g\in \mathbb{R}$ and $c_{i},c_{\mu },c_{\nu }=\pm 1$. While
this action $\mathcal{I}_{3}$ is not Hermitian, that is invariant under
complex conjugation, it respects various discrete and continuous symmetries.
It is invariant under two types of $\mathcal{CPT}$-transformations (\ref%
{CPT12})%
\begin{equation}
\mathcal{CPT}_{1/2}:\phi _{1}(x_{\mu })\rightarrow \pm \phi _{1}^{\ast
}(-x_{\mu })\,\text{,\quad\ }\phi _{2}(x_{\mu })\rightarrow \mp \phi
_{2}^{\ast }(-x_{\mu })\text{, \quad }\phi _{3}(x_{\mu })\rightarrow \pm
\phi _{3}^{\ast }(-x_{\mu })\text{,}  \label{cpt1}
\end{equation}%
which are both discrete antilinear transformations. Moreover, the action (%
\ref{I1}) is left invariant under the continuous global $U(1)$-symmetry (\ref%
{U11}), which gives rise to the Noether current (\ref{JN}) 
\begin{equation}
j_{\mu }=i\alpha \sum\nolimits_{i=1}^{3}\left( \phi _{i}\partial _{\mu }\phi
_{i}^{\ast }-\phi _{i}^{\ast }\partial _{\mu }\phi _{i}\right) .
\end{equation}%
With the dimension of the global symmetry group $G=U(1)$ being just $1$, we
may only encounter two possibilities for the Hermitian case, that is the
model contains one or no massless Goldstone boson when the symmetry group
for the expanded theory is $H=\mathbb{I}$ or $H=U(1)$, respectively, after a
specific vacuum has been selected \cite%
{nambu1961dynamical,goldstone1961field}. As we shall see, breaking in our
model the global $U(1)$-symmetry for the vacuum will give rise to the
massless Goldstone bosons in the standard fashion, albeit with some
modifications and novel features for a non-Hermitian setting. The six
equations of motion in (\ref{EoM}) read in this case%
\begin{eqnarray}
\square \phi _{1}-c_{1}m_{1}^{2}\phi _{1}-c_{\mu }\mu ^{2}\phi _{2}+\frac{g}{%
2}\phi _{1}^{2}\phi _{1}^{\ast } &=&0,  \label{v1} \\
\square \phi _{2}-c_{2}m_{2}^{2}\phi _{2}+c_{\mu }\mu ^{2}\phi _{1}+c_{\nu
}\nu ^{2}\phi _{3} &=&0, \\
\square \phi _{3}-c_{3}m_{3}^{2}\phi _{3}-c_{\nu }\nu ^{2}\phi _{2} &=&0,
\label{v3} \\
\square \phi _{1}^{\ast }-c_{1}m_{1}^{2}\phi _{1}^{\ast }+c_{\mu }\mu
^{2}\phi _{2}^{\ast }+\frac{g}{2}\phi _{1}(\phi _{1}^{\ast })^{2} &=&0,
\label{v4} \\
\square \phi _{2}^{\ast }-c_{2}m_{2}^{2}\phi _{2}^{\ast }-c_{\mu }\mu
^{2}\phi _{1}^{\ast }-c_{\nu }\nu ^{2}\phi _{3}^{\ast } &=&0, \\
\square \phi _{3}^{\ast }-c_{3}m_{3}^{2}\phi _{3}^{\ast }+c_{\nu }\nu
^{2}\phi _{2}^{\ast } &=&0,  \label{v6}
\end{eqnarray}%
with d'Alembert operator $\square :=\partial _{\mu }\partial ^{\mu }$ and
metric $\limfunc{diag}\eta =(1,-1,-1,-1)$. We encounter here the same
problem as pointed out for $\mathcal{I}_{2}$ with four scalar fields
investigated in \cite{alexandre2018spontaneous,mannheim2018goldstone},
namely that as a consequence of the non-Hermiticity of the action the
equations of motions obtained from the variation with regard to the fields $%
\phi _{i}^{\ast }$, (\ref{v1})-(\ref{v3}), are not the complex conjugates of
the equations obtained from the variation with respect to the fields $\phi
_{i}$, (\ref{v4})-(\ref{v6}). Hence, the two sets of equations appear to be
incompatible and therefore the quantum field theory related to the action (%
\ref{I1}) seems to be inconsistent.

An unconventional solution to this conundrum was proposed in \cite%
{alexandre2018spontaneous}, by suggesting to omit the variation with respect
to one set of fields and also taking non-vanishing surface terms into
account. Even though this proposal appears to lead to a consistent model, it
remains somewhat unclear as to why one should abandon a well established
principle from standard complex scalar field theory. Here we adopt the
proposal made by Mannheim \cite{mannheim2018goldstone}, which is more
elegant and, from the point of view of extending the well established
framework of non-Hermitian quantum mechanics to quantum field theory, also
more natural. It consists of seeking a similarity transformation for the
action that achieves compatibility between the two sets of equations of
motion. It is easy to see that any transformation of the form $\phi
_{2}\rightarrow \pm i\phi _{2}$, $\phi _{2}^{\ast }\rightarrow \pm i\phi
_{2}^{\ast }~$that leaves all the other fields invariant will achieve
compatibility between the two sets of equations (\ref{v1})-(\ref{v3}) and (%
\ref{v4})-(\ref{v6})$.$

The analysis to achieve this is most conveniently carried out when
reparameterising the complex fields in terms of real component fields.
Parameterising therefore the complex scalar field as $\phi _{i}=1/\sqrt{2}%
(\varphi _{i}+i\chi _{i})$ with $\varphi _{i}$, $\chi _{i}\in \mathbb{R}$
the action $\mathcal{I}_{3}$ in (\ref{I1}) acquires the form 
\begin{eqnarray}
\mathcal{I}_{3} &=&\int d^{4}x\left\{ \sum\limits_{i=1}^{3}\frac{1}{2}\left[
\partial _{\mu }\varphi _{i}\partial ^{\mu }\varphi _{i}+\partial _{\mu
}\chi _{i}\partial ^{\mu }\chi _{i}+c_{i}m_{i}^{2}\left( \varphi
_{i}^{2}+\chi _{i}^{2}\right) \right] +ic_{\mu }\mu ^{2}\left( \varphi
_{1}\chi _{2}-\varphi _{2}\chi _{1}\right) \right. ~~~\ \ ~~~  \label{I3F} \\
&&\left. +ic_{\nu }\nu ^{2}\left( \varphi _{3}\chi _{2}-\varphi _{2}\chi
_{3}\right) -\frac{g}{16}(\varphi _{1}^{2}+\chi _{1}^{2})^{2}\right\} . 
\notag
\end{eqnarray}%
This approach differs slightly from Mannheim's, who took the component
fields to be complex as well. The continuous global $U(1)$-symmetry (\ref%
{U11}) of the action is realised for the real fields as $\varphi
_{i}\rightarrow \varphi _{i}\cos \alpha -\chi _{i}\sin \alpha $, $\chi
_{i}\rightarrow \varphi _{i}\sin \alpha +\chi _{i}\cos \alpha $, that is $%
\delta \varphi _{i}=-\alpha \chi _{i}$ and $\delta \chi _{i}=\alpha \varphi
_{i}$ for $\alpha $ small. The $\mathcal{CPT}_{1/2}$ symmetries in (\ref%
{cpt1}) manifests on these fields as 
\begin{eqnarray}
\mathcal{CPT}_{1/2} &:&\varphi _{1,3}(x_{\mu })\rightarrow \pm \varphi
_{1,3}(-x_{\mu })\,\text{,\quad }\varphi _{2}(x_{\mu })\rightarrow \mp
\varphi _{2}(-x_{\mu })\text{,\quad }  \label{CPT2} \\
&&\chi _{1,3}(x_{\mu })\rightarrow \pm \chi _{1,3}(-x_{\mu })\text{,\quad }%
\chi _{2}(x_{\mu })\rightarrow \mp \chi _{2}(-x_{\mu })\text{,\quad }%
i\rightarrow -i\text{.}  \notag
\end{eqnarray}%
In this form also the antilinear symmetry%
\begin{equation*}
\mathcal{CPT}_{3/4}:\varphi _{1,2,3}(x_{\mu })\rightarrow \pm \chi
_{1,2,3}(-x_{\mu })\,\text{,\quad }\chi _{1,2,3}(x_{\mu })\rightarrow \pm
\varphi _{1,2,3}(-x_{\mu })\text{,\quad }i\rightarrow -i\text{,}
\end{equation*}%
leaves the action invariant. Let us now transform the action $\mathcal{I}%
_{3} $ in the form (\ref{I3F}) to an equivalent Hermitian one.

\subsection{A $\mathcal{CPT}$ equivalent action, different types of vacua}

We define now the analogue to the Dyson map \cite{Dyson} in quantum
mechanics as%
\begin{equation}
\eta =\exp \left[ \frac{\pi }{2}\int d^{3}x\Pi _{2}^{\varphi }(\mathbf{x}%
,t)\varphi _{2}(\mathbf{x},t)\right] \exp \left[ \frac{\pi }{2}\int
d^{3}x\Pi _{2}^{\chi }(\mathbf{x},t)\chi _{2}(\mathbf{x},t)\right] ,
\label{eta}
\end{equation}%
involving the canonical momenta $\Pi _{i}^{\varphi }=\partial _{t}\varphi
_{i}$ and $\Pi _{i}^{\chi }=\partial _{t}\chi _{i}$, $i=1,2,3$. Using the
Baker-Campbell-Haussdorf formula we compute the adjoint actions of $\eta $
on the scalar fields as%
\begin{equation}
\eta \varphi _{i}\eta ^{-1}=(-i)^{\delta _{2i}}\varphi _{i},~~~\eta \chi
_{i}\eta ^{-1}=(-i)^{\delta _{2i}}\chi _{i},~~~~~\eta \phi _{i}\eta
^{-1}=(-i)^{\delta _{2i}}\phi _{i},~~~~\eta \phi _{i}^{\ast }\eta
^{-1}=(-i)^{\delta _{2i}}\phi _{i}^{\ast }.
\end{equation}%
The equal time commutation relations $\left[ \psi _{j}(\mathbf{x},t),\Pi
_{j}^{\psi _{j}}(\mathbf{y},t)\right] =i\delta (\mathbf{x}-\mathbf{y})$, $%
i=1,2,3$, for $\psi =\varphi ,\chi $ are preserved under these
transformations. Applying them to $\mathcal{I}_{3}$ in (\ref{I3F}), we
obtain the new equivalent action 
\begin{eqnarray}
\hat{{\mathcal{I}}}_{3}=\eta \mathcal{I}_{3}\eta ^{-1}= &&\int
d^{4}x\sum\limits_{i=1}^{3}\frac{1}{2}(-1)^{\delta _{2i}}\left[ \partial
_{\mu }\varphi _{i}\partial ^{\mu }\varphi _{i}+\partial _{\mu }\chi
_{i}\partial ^{\mu }\chi _{i}+c_{i}m_{i}^{2}\left( \varphi _{i}^{2}+\chi
_{i}^{2}\right) \right] ~~~\ \ ~ \\
&&+c_{\mu }\mu ^{2}\left( \varphi _{1}\chi _{2}-\varphi _{2}\chi _{1}\right)
+c_{\nu }\nu ^{2}\left( \varphi _{3}\chi _{2}-\varphi _{2}\chi _{3}\right) -%
\frac{g}{16}(\varphi _{1}^{2}+\chi _{1}^{2})^{2}.  \notag
\end{eqnarray}%
The $U(1)$-symmetry is still realised in the same way as for $\mathcal{I}%
_{3} $, but the $\mathcal{CPT}$-symmetries for $\hat{{\mathcal{I}}}_{3}$ are
now modified to 
\begin{eqnarray}
\widehat{\mathcal{CPT}}_{1/2} &:&\varphi _{1,3}(x_{\mu })\rightarrow \pm
\varphi _{1,3}(-x_{\mu })\,\text{,\quad }\varphi _{2}(x_{\mu })\rightarrow
\mp \varphi _{2}(-x_{\mu })\text{,\quad }  \label{cpthat} \\
&&\chi _{1,3}(x_{\mu })\rightarrow \mp \chi _{1,3}(-x_{\mu })\text{,\quad }%
\chi _{2}(x_{\mu })\rightarrow \pm \chi _{2}(-x_{\mu })\text{,}  \notag \\
\widehat{\mathcal{CPT}}_{3/4} &:&\varphi _{1,2,3}(x_{\mu })\rightarrow \pm
\chi _{1,2,3}(-x_{\mu })\,\text{,\quad }
\end{eqnarray}%
accommodating the fact that no explicit imaginary unit $i$ is left in the
action. Notice that these symmetries are, however, no longer antilinear and
therefore lack the constraining power of predicting the reality of
non-Hermitian quantities. The equations of motion resulting from
functionally varying $\hat{{\mathcal{I}}}_{3}$ with respect to the real
fields are 
\begin{eqnarray}
-\square \varphi _{1} &=&\frac{\partial V}{\partial \varphi _{1}}%
=-c_{1}m_{1}^{2}\varphi _{1}-c_{\mu }\mu ^{2}\chi _{2}+\frac{g}{4}\varphi
_{1}(\varphi _{1}^{2}+\chi _{1}^{2}),  \label{eq1} \\
-\square \chi _{2} &=&-\frac{\partial V}{\partial \chi _{2}}%
=-c_{2}m_{2}^{2}\chi _{2}+c_{\mu }\mu ^{2}\varphi _{1}+c_{\nu }\nu
^{2}\varphi _{3},  \label{eq2} \\
-\square \varphi _{3} &=&\frac{\partial V}{\partial \varphi _{3}}%
=-c_{3}m_{3}^{2}\varphi _{3}-c_{\nu }\nu ^{2}\chi _{2}, \\
-\square \chi _{1} &=&\frac{\partial V}{\partial \chi _{1}}%
=-c_{1}m_{1}^{2}\chi _{1}+c_{\mu }\mu ^{2}\varphi _{2}+\frac{g}{4}\chi
_{1}(\varphi _{1}^{2}+\chi _{1}^{2}),  \label{eq4} \\
-\square \varphi _{2} &=&-\frac{\partial V}{\partial \varphi _{2}}%
=-c_{2}m_{2}^{2}\varphi _{2}-c_{\mu }\mu ^{2}\chi _{1}-c_{\nu }\nu ^{2}\chi
_{3},  \label{eq5} \\
-\square \chi _{3} &=&\frac{\partial V}{\partial \chi _{3}}%
=-c_{3}m_{3}^{2}\chi _{3}+c_{\nu }\nu ^{2}\varphi _{2}.  \label{eq6}
\end{eqnarray}%
We may write the action $\hat{{\mathcal{I}}}_{3}$ and the corresponding
equation of motions more compactly. Introducing the column vector field $%
\Phi =(\varphi _{1},\chi _{2},\varphi _{3},\chi _{1},\varphi _{2},\chi
_{3})^{T}$, the action acquires the concise form 
\begin{equation}
\hat{{\mathcal{I}}}_{3}=\frac{1}{2}\int d^{4}x\left[ \partial _{\mu }\Phi
^{T}I\partial ^{\mu }\Phi -\Phi ^{T}H_{t}\Phi -\frac{g}{8}\left( \Phi
^{T}E\Phi \right) ^{2}\right] .  \label{I3}
\end{equation}%
Here we employed the Hessian matrix $H_{ij}(\Phi )=\left. \frac{\partial
^{2}V}{\partial \Phi _{i}\partial \Phi _{j}}\right\vert _{\Phi }$ which for
our potential $V_{3}$ reads 
\begin{equation}
H\left( \Phi \right) =\left( 
\begin{array}{cccccc}
\frac{g}{4}(3\varphi _{1}^{2}+\chi _{1}^{2})-c_{1}m_{1}^{2} & -c_{\mu }\mu
^{2} & 0 & \frac{g}{2}\varphi _{1}\chi _{1} & 0 & 0 \\ 
-c_{\mu }\mu ^{2} & c_{2}m_{2}^{2} & -c_{\nu }\nu ^{2} & 0 & 0 & 0 \\ 
0 & -c_{\nu }\nu ^{2} & -c_{3}m_{3}^{2} & 0 & 0 & 0 \\ 
\frac{g}{2}\varphi _{1}\chi _{1} & 0 & 0 & \frac{g}{4}(\varphi
_{1}^{2}+3\chi _{1}^{2})-c_{1}m_{1}^{2} & c_{\mu }\mu ^{2} & 0 \\ 
0 & 0 & 0 & c_{\mu }\mu ^{2} & c_{2}m_{2}^{2} & c_{\nu }\nu ^{2} \\ 
0 & 0 & 0 & 0 & c_{\nu }\nu ^{2} & -c_{3}m_{3}^{2}%
\end{array}%
\right) .
\end{equation}%
In (\ref{I3}) we use $H_{t}=H\left( \Phi _{1}^{0}\right) $, $\Phi
_{1}^{0}=(0,0,0,0,0,0)$ and the $6\times 6$-matrices $I$, $E$ with $\limfunc{%
diag}I=(1,-1,1,1,-1,1)$ and $\limfunc{diag}E=(1,0,0,1,0,0)$. The equation of
motion resulting from (\ref{I3}) reads%
\begin{equation}
-\square \Phi -IH_{t}\Phi -\frac{g}{4}I\left( \Phi ^{T}E\Phi \right) E\Phi
=0.  \label{EQ}
\end{equation}%
We find different types of vacua by solving $\delta V=0$, amounting to
setting simultaneously the right hand sides of the equations (\ref{eq1})-(%
\ref{eq6}) to zero and solving for the fields\ $\varphi _{i},\chi _{i}$.
Denoting the solutions by $\Phi ^{0}=(\varphi _{1}^{0},\chi _{2}^{0},\varphi
_{3}^{0},\chi _{1}^{0},\varphi _{2}^{0},\chi _{3}^{0})^{T}$, we find the
vacua%
\begin{eqnarray}
\Phi _{1}^{0} &=&(0,0,0,0,0,0),  \label{vac1} \\
\Phi _{2}^{0} &=&K(0)\left( 1,\frac{c_{3}c_{\mu }m_{3}^{2}\mu ^{2}}{\kappa }%
,-\frac{c_{3}c_{\mu }m_{3}^{2}\mu ^{2}}{\kappa },0,0,0\right) , \\
\Phi _{3}^{0} &=&K(0)\left( 0,0,0,-1,\frac{c_{3}c_{\mu }m_{3}^{2}\mu ^{2}}{%
\kappa },\frac{c_{\nu }c_{\mu }\nu ^{2}\mu ^{2}}{\kappa }\right) , \\
\Phi _{4}^{0} &=&\left( \varphi _{1}^{0},\frac{c_{3}c_{\mu }m_{3}^{2}\mu
^{2}\varphi _{1}^{0}}{\kappa },-\frac{c_{\nu }c_{\mu }\nu ^{2}\mu
^{2}\varphi _{1}^{0}}{\kappa },-K(\varphi _{1}^{0}),\frac{c_{3}c_{\mu
}m_{3}^{2}\mu ^{2}K(\varphi _{1}^{0})}{\kappa },\frac{c_{\nu }c_{\mu }\nu
^{2}\mu ^{2}K(\varphi _{1}^{0})}{\kappa }\right) ,~~~~~~~  \label{vac4}
\end{eqnarray}%
where for convenience we introduced the function and constant 
\begin{equation}
K(x):=\pm \sqrt{\frac{4c_{3}m_{3}^{2}\mu ^{4}}{g\kappa }+\frac{%
4c_{1}m_{1}^{2}}{g}-x^{2}},\quad \text{\quad }\kappa
:=c_{2}c_{3}m_{2}^{2}m_{3}^{2}+\nu ^{4}.  \label{Kk}
\end{equation}%
Notice, that in the vacuum $\Phi _{4}^{0}$ the field $\varphi _{1}^{0}$ is
generic and not fixed. When varied it interpolates between the vacua $\Phi
_{2}^{0}$ and $\Phi _{3}^{0}$. For $(\varphi _{1}^{0})^{2}\rightarrow
4(c_{1}m_{1}^{2}\kappa +c_{3}m_{3}^{2}\mu ^{4})/g\kappa $ and $\varphi
_{1}^{0}\rightarrow 0$ we obtain $\Phi _{4}^{0}\rightarrow \Phi _{2}^{0}$
and $\Phi _{4}^{0}\rightarrow \Phi _{3}^{0}$, respectively. We also note
that $K(0)=$ $0$ at the special value of the coupling $\mu =\mu
_{s}^{4}=-c_{1}m_{1}^{2}\kappa /c_{3}m_{3}^{2}$ so that $\Phi _{2}^{0}(\mu
_{s})=\Phi _{1}^{0}$. Next we probe Goldstone's theorem by computing the
masses resulting by expanding around the different vacua in the tree
approximation.

\subsection{The mass spectra, $\mathcal{PT}$-symmetries}

Defining the column vector field $\Phi =\Phi ^{0}+\hat{\Phi}$ with vacuum
component $\Phi ^{0}$ as defined above and $\hat{\Phi}=(\hat{\varphi}_{1},%
\hat{\chi}_{2},\hat{\varphi}_{3},\hat{\chi}_{1},\hat{\varphi}_{2},\hat{\chi}%
_{3})^{T}$, we expand the potential about the vacua (\ref{vac1})-(\ref{vac4}%
) as%
\begin{equation}
V\left( \Phi \right) =V\left( \Phi ^{0}+\hat{\Phi}\right) =V\left( \Phi
^{0}\right) +\nabla V\left( \Phi ^{0}\right) ^{T}\hat{\Phi}+\frac{1}{2}\hat{%
\Phi}^{T}H\left( \Phi ^{0}\right) \hat{\Phi}+\ldots .
\end{equation}%
The linear term is of course vanishing, as by design $\nabla V\left( \Phi
^{0}\right) =0$. The squared mass matrix $M^{2}$ is read off from (\ref{EQ})
as 
\begin{equation}
\left( M^{2}\right) _{ij}=[IH\left( \Phi ^{0}\right) ]_{ij}\text{.}
\label{MH}
\end{equation}%
The somewhat unusual emergence of the matrix $I$ is due to the fact that as
a consequence of the similarity transformation we now have negative signs in
front of some of the kinetic energy terms, see also (\ref{eq2}) and (\ref%
{eq5}).

In general this matrix is not diagonal, but in the $\mathcal{CPT}$-symmetric
regime we may diagonalise it and express the fields related to these masses
in terms of the original fields in the action. Denoting the eigenvectors of
the squared mass matrix by $v_{i}$, $i=1,\ldots ,6$, the matrix $%
U=(v_{1},\ldots ,v_{6})$, containing the eigenvectors as column vectors,
diagonalizes $M^{2}$ as $U^{-1}M^{2}U=D$ with $\limfunc{diag}D=(\lambda
_{1},\ldots ,\lambda _{6})$ as long as $U$ is invertible. The latter
property holds in general only in the $\mathcal{CPT}$-symmetric regime.
Rewriting%
\begin{equation}
\hat{\Phi}^{T}M^{2}\hat{\Phi}=\sum\nolimits_{i}m_{i}^{2}\psi
_{i}^{2}=\sum\nolimits_{i}m_{i}^{2}\left( \hat{\Phi}^{T}IU\right) _{i}(U^{-1}%
\hat{\Phi})_{i},  \label{psim}
\end{equation}%
we may therefore introduce the masses $m_{i}$ for the fields 
\begin{equation}
\psi _{i}:=\sqrt{\left( \hat{\Phi}^{T}IU\right) _{i}(U^{-1}\hat{\Phi})_{i}}
\label{psi}
\end{equation}%
as the positive square roots of the eigenvalues of the squared mass matrix $%
M^{2}$, that is $m_{i}=\sqrt{\lambda _{i}}$. Naturally this means the fields 
$\psi _{i}$ in the specific form (\ref{psi}) are absent when $U$ is not
invertible and since physical masses $m_{i}$ are non-negative we must also
discard scenarios in which $\lambda _{i}<0$ or $\func{Im}\lambda _{i}\neq 0$
as unphysical.

Since the squared mass matrix $M^{2}$ is not Hermitian, but may have real
eigenvalues $\lambda _{i}$ in some regime, we can employ the standard
framework from $\mathcal{PT}$-symmetric quantum mechanics with $M^{2}$
playing the role of the non-Hermitian Hamiltonian \cite{Bender:1998ke,PTbook}%
. We can then identify the antilinear $\mathcal{PT}$-operator that ensures
the reality of the spectrum in that particular regime. The time-reversal
operator $\mathcal{T}$\ simply corresponds to a complex conjugation, but one
needs to establish that the $\mathcal{P}$-operator obtained from the quantum
mechanical description is the same as the one employed at the level of the
action. In order to identify that connection let us first see which
properties the $\mathcal{P}$-operator must satisfy at the level of the
action. Expressing $\mathcal{I}_{3}$ in the form 
\begin{equation}
\mathcal{I}_{3}\left[ \Phi \right] =\mathcal{I}_{3}^{\text{M}}\left[ \Phi %
\right] +\mathcal{I}_{3}^{\text{int}}\left[ \Phi \right] =\frac{1}{2}\int
d^{4}x\left[ \Phi ^{T}\left( \square +M^{2}\right) \Phi \right] +\mathcal{I}%
_{3}^{\text{int}}\left[ \Phi \right] .
\end{equation}%
with real field vector $\Phi $, the action of the $\mathcal{CPT}$-operator
on $\mathcal{I}_{3}^{\text{M}}\left[ \Phi \right] $ is%
\begin{equation}
\mathcal{CPT}:\mathcal{I}_{3}^{\text{M}}\left[ \Phi \right] \rightarrow 
\frac{1}{2}\int d^{4}x\left[ \Phi ^{T}\left[ \mathcal{P}^{T}\mathcal{P}%
\square +\mathcal{P}^{T}\left( M^{2}\right) ^{\ast }\mathcal{P}\right] \Phi %
\right] .
\end{equation}%
Hence for this part of the action to be invariant we require the $\mathcal{P}
$-operator to obey the two relations%
\begin{equation}
\mathcal{P}^{T}\mathcal{P=}\mathbb{I},\quad \text{and\quad }\left(
M^{2}\right) ^{\ast }\mathcal{P}=\mathcal{P}M^{2}.  \label{PTM}
\end{equation}%
This is in fact the same property $\mathcal{P}$ needs to satisfy in the $%
\mathcal{PT}$-quantum mechanical framework. Let us see how to construct $%
\mathcal{P}$ when given the non-Hermitian matrix $M^{2}$. We start by
constructing a biorthonormal basis from the left and right eigenvectors $%
u_{n}$ and $v_{n}$, respectively, of $M^{2}$%
\begin{equation}
M^{2}v_{n}=\varepsilon _{n}v_{n},\qquad \left( M^{2}\right) ^{\dagger
}u_{n}=\varepsilon _{n}u_{n}  \label{leftright}
\end{equation}%
satisfying%
\begin{equation}
\left\langle u_{n}\right. \left\vert v_{n}\right\rangle =\delta _{nm},\qquad
\sum\nolimits_{n}\left\vert u_{n}\right\rangle \left\langle v_{n}\right\vert
=\sum\nolimits_{n}\left\vert v_{n}\right\rangle \left\langle
u_{n}\right\vert =\mathbb{I}\text{.}  \label{uv}
\end{equation}%
The left and right eigenvectors are related by the $\mathcal{P}$-operator as 
\begin{equation}
\left\vert u_{n}\right\rangle =s_{n}\mathcal{P}\left\vert v_{n}\right\rangle
.  \label{lr}
\end{equation}%
with $s_{n}=\pm 1$ defining the signature. Combining (\ref{lr}), (\ref{uv})
and the first relation in (\ref{PTM}) we can express the $\mathcal{P}$%
-operator and its transpose in terms of the left and right eigenvectors as%
\begin{equation}
\mathcal{P=}\sum\nolimits_{n}s_{n}\left\vert u_{n}\right\rangle \left\langle
u_{n}\right\vert ,\qquad \text{and\qquad }\mathcal{P}^{T}\mathcal{=}%
\sum\nolimits_{n}s_{n}\left\vert v_{n}\right\rangle \left\langle
v_{n}\right\vert .  \label{PPT}
\end{equation}%
The biorthonormal basis can also be used to construct an operator, often
denoted with the symbol $C$, that is closely related to the metric $\rho $
used in non-Hermitian quantum mechanics 
\begin{equation}
C=\mathcal{P}^{T}\rho =\sum\nolimits_{n}s_{n}\left\vert v_{n}\right\rangle
\left\langle u_{n}\right\vert .  \label{C}
\end{equation}%
Despite its notation, this operator is not to be confused with the charge
conjugation operator $\mathcal{C}$ employed on the level of the action. The
operator $C$ satisfies the algebraic properties \cite{bender2002complex}%
\begin{equation}
\left[ C,M^{2}\right] =0,\qquad \left[ C,\mathcal{PT}\right] =0,\qquad C^{2}=%
\mathbb{I}.  \label{Calg}
\end{equation}%
When compared to the quantum mechanical setting the operator $U^{-1}$ plays
here the analogue to the Dyson map $\eta $ and the combination $\left(
U^{-1}\right) ^{\dagger }U^{-1}$ is the analogue to the metric operator $%
\rho $. However, constructing $\mathcal{P}$ with $M^{2}$ as a starting point
does of course not guarantee that also $\mathcal{I}_{3}^{\text{int}}\left[
\Phi \right] $ will be invariant under $\mathcal{CPT}$ when using this
particular $\mathcal{P}$-operator. In fact, we shall see below that there
are many solutions to the two relations in (\ref{PTM}) that do not leave $%
\mathcal{I}_{3}^{\text{int}}\left[ \Phi \right] $ invariant. Thus for these $%
\mathcal{CPT}$ -operators the symmetry is broken on the level of the action,
but the mass spectra would still be real as the symmetry is preserved at the
tree approximation and the breaking only occurs at higher order.

\subsection{$U(1)$ and $\mathcal{CPT}$ invariant vacuum, absence of
Goldstone bosons}

We investigate now in more detail the theory expanded about the vacuum $\Phi
_{1}^{0}$ in (\ref{vac1}). According to our discussion at the end of the
last section the theory expanded about this vacuum is\ invariant under the
global $U(1)$-symmetry and all four $\mathcal{CPT}$-symmetries. As the
dimension of the coset $G/H$ equals $0$ the standard field theoretical
arguments on Goldstone's theorem suggest that we do not expect a Goldstone
boson to emerge when expanding around this vacuum. We confirm this by
considering the squared mass matrix as defined in (\ref{MH}), which for this
vacuum decomposes into Jordan block form as%
\begin{equation}
M_{1}^{2}=\left( 
\begin{array}{cccccc}
-c_{1}m_{1}^{2} & -c_{\mu }\mu ^{2} & 0 & 0 & 0 & 0 \\ 
c_{\mu }\mu ^{2} & -c_{2}m_{2}^{2} & c_{\nu }\nu ^{2} & 0 & 0 & 0 \\ 
0 & -c_{\nu }\nu ^{2} & -c_{3}m_{3}^{2} & 0 & 0 & 0 \\ 
0 & 0 & 0 & -c_{1}m_{1}^{2} & c_{\mu }\mu ^{2} & 0 \\ 
0 & 0 & 0 & -c_{\mu }\mu ^{2} & -c_{2}m_{2}^{2} & -c_{\nu }\nu ^{2} \\ 
0 & 0 & 0 & 0 & c_{\nu }\nu ^{2} & -c_{3}m_{3}^{2}%
\end{array}%
\right) ,
\end{equation}%
where we label the entries of the matrix by the fields in the order as
defined for the vector field $\Phi $. The two blocks are simply related as $%
c_{\nu /\mu }\rightarrow -c_{\nu /\mu }$. We find that the eigenvalues of
each block only depend on the combination $c_{\nu /\mu }^{2}=1$, so that we
have three degenerate eigenvalues with linear independent eigenvectors and
it therefore suffices to consider one block only and subsequently implement
the degeneracy. Evaluating the constant term of the third order
characteristic equations we obtain $-c_{3}m_{3}^{2}\mu
^{4}-c_{1}m_{1}^{2}\nu ^{4}-c_{1}c_{2}c_{3}m_{1}^{2}m_{2}^{2}m_{3}^{2}$ for
each block. In general, this is not equal to zero indicating the absence of
a massless Goldstone boson as expected or any other type of massless
particle. The two choices $c_{1}=c_{2}=c_{3}=\pm 1$ exclude the possibility
for this term to vanish for \textit{any} values in parameter space $%
(m_{1},m_{2},m_{3},\mu ,\nu )$. Alternatively this is also seen from $\det
M_{1}^{2}=(c_{3}m_{3}^{2}\mu ^{4}+c_{1}m_{1}^{2}\kappa )^{2}$ with $\kappa $
as defined in (\ref{Kk}).

All other choices for the constants $c_{i}$ may lead to zero masses for
specific values in the parameters space. For instance, when $%
c_{1}=-c_{2}=c_{3}=1$, the linear term vanishes for the special choice $\mu
_{s}=(m_{1}^{2}m_{2}^{2}-\nu ^{4}m_{1}^{2}/m_{3}^{2})^{1/4}$, so that we
obtain two zero mass particles in the spectrum, of which, however, none is a
Goldstone boson. As in the general case with unrestricted $\mu $, the
eigenvalues $\lambda $ of $M_{1}^{2}$ indicate some unphysical regions, with 
$\lambda $ being either negative or complex. However, the model has also a
physical region in which two degenerate eigenvalues of the squared mass
matrix are positive and, somewhat unexpectedly from the symmetry argument,
there are also two massless particles present in the spectrum. The behaviour
of the remaining two degenerate eigenvalues is depicted in figure \ref{Fig1}.

\FIGURE{ \epsfig{file=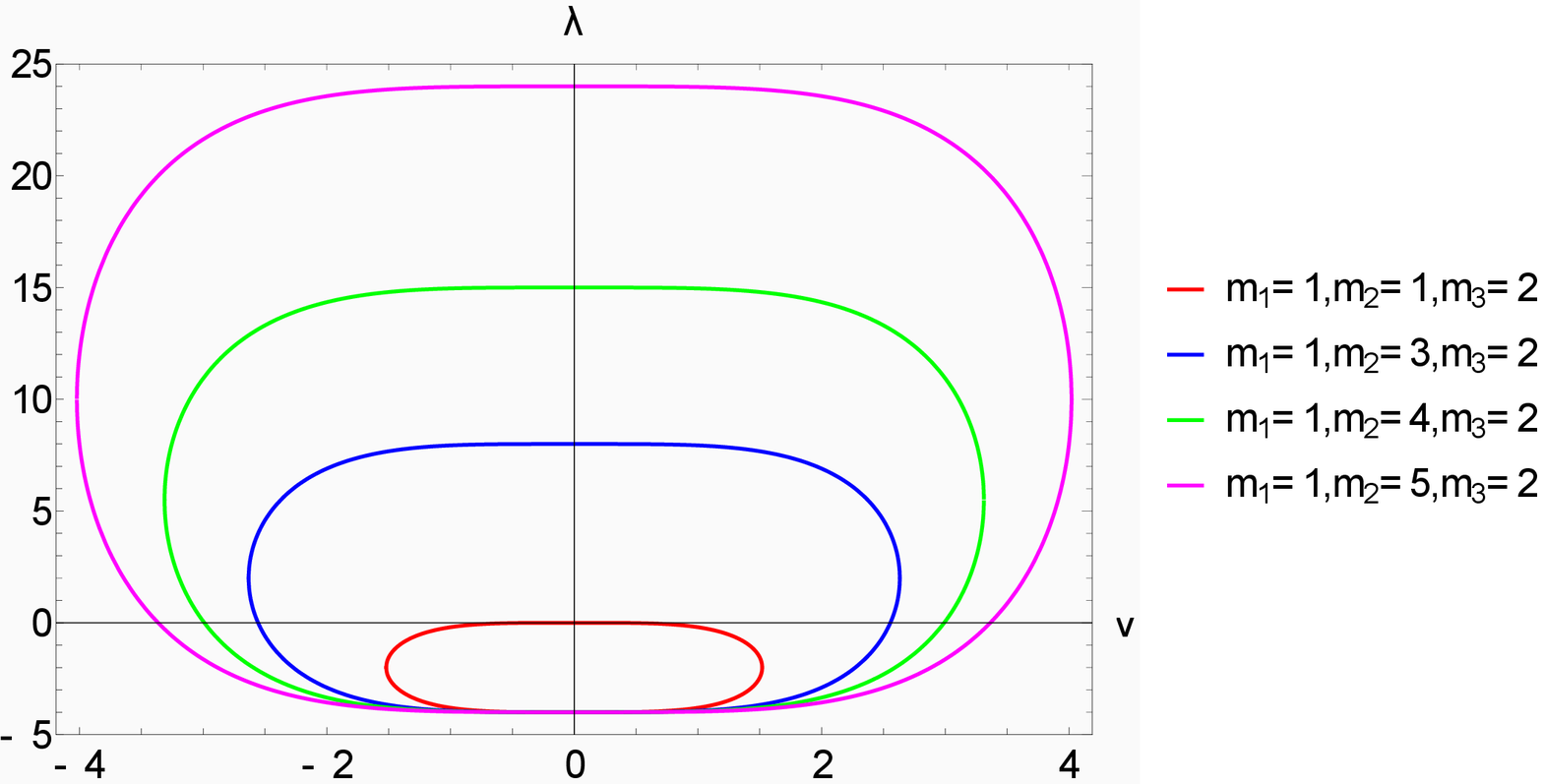,width=11.0cm} 
\caption{Nonvanishing eigenvalues $\lambda$ of  $M_{1}^{2}$ as a function
of $\nu $ for $c_{1}=-c_{2}=c_{3}=1$, at the special point $\mu =\mu_s$, fixed values for $m_{1}$ and $m_{3}$ at different
values of $m_{2}$. The physical regions are $\nu \in (-\nu _{\text{ex}},-\nu _{0})$ and $\nu \in (\nu _{0},\nu _{\text{ex}})$.}
        \label{Fig1}}

The region $\nu \in (-\nu _{0},\nu _{0})$ with $\nu
_{0}=m_{3}(m_{1}^{2}-m_{2}^{2})^{1/4}/(m_{1}^{2}-m_{3}^{2})^{1/4}$ is
therefore discarded as unphysical because one of the eigenvalues of $%
M_{1}^{2}$ is negative. At $\pm \nu _{\text{ex}}$, with $\nu _{\text{ex}%
}=[m_{3}^{2}(m_{2}^{2}+m_{3}^{2}-m_{1}^{2})^{2}/4/(m_{3}^{2}-m_{1}^{2})]^{1/4} 
$ for $m_{3}^{2}>m_{1}^{2}$, the two eigenvalues coalesce and become a
complex conjugate pair, a scenario that for the energy spectrum in the
quantum mechanical context is usually referred to as an exceptional point.
Hence, also the regions $\nu <-\nu _{\text{ex}}$ and $\nu >\nu _{\text{ex}}$
are excluded as being unphysical. Crucially, however, the model is not empty
and possess a physical region in parameter space.

\subsection{$U(1)$ broken and $\mathcal{CPT}$-invariant vacua, presence of
Goldstone bosons}

Let us next choose another vacuum that breaks the global $U(1)$-symmetry. In
this case we expect one massless Goldstone boson to appear. However, as in
the previous case there are some regions in the parameter space for which
the model may possess a second massless particle. We choose now the vacuum $%
\Phi _{2}^{0}$. Notice that for $c_{1}=-c_{2}=c_{3}=1$ and $\mu \rightarrow
\mu _{s}$, as defined above, the global symmetry breaking and symmetry
preserving vacua coincide $\Phi _{2}^{0}\rightarrow \Phi _{1}^{0}$, and
therefore the previous discussion applies in that case. Expanding the action
around this $U(1)$-symmetry breaking vacuum for $\mu \neq \mu _{s}$, the
corresponding squared mass matrix becomes%
\begin{equation}
M_{2}^{2}=\left( 
\begin{array}{cccccc}
\frac{3c_{3}m_{3}^{2}\mu ^{4}}{\kappa }+2c_{1}m_{1}^{2} & -c_{\mu }\mu ^{2}
& 0 & 0 & 0 & 0 \\ 
c_{\mu }\mu ^{2} & -c_{2}m_{2}^{2} & c_{\nu }\nu ^{2} & 0 & 0 & 0 \\ 
0 & -c_{\nu }\nu ^{2} & -c_{3}m_{3}^{2} & 0 & 0 & 0 \\ 
0 & 0 & 0 & \frac{c_{3}m_{3}^{2}\mu ^{4}}{\kappa } & c_{\mu }\mu ^{2} & 0 \\ 
0 & 0 & 0 & -c_{\mu }\mu ^{2} & -c_{2}m_{2}^{2} & -c_{\nu }\nu ^{2} \\ 
0 & 0 & 0 & 0 & c_{\nu }\nu ^{2} & -c_{3}m_{3}^{2}%
\end{array}%
\right) ,  \label{mv2}
\end{equation}%
with $\det M_{2}^{2}=0$, hence indicating a zero eigenvalue. Let us now
comment on where this Goldstone boson originates from. Both blocks in $%
M_{2}^{2}$ are of the following general $3\times 3$-matrix form%
\begin{equation}
\left( 
\begin{array}{ccc}
A & W & 0 \\ 
-W & B & -V \\ 
0 & V & -C%
\end{array}%
\right) ,
\end{equation}%
whose eigenvalues are solutions to the cubic characteristic equation $%
\lambda ^{3}+r\lambda ^{2}+s\lambda +t=0$ with 
\begin{equation}
r=C-A-B,~~s=V^{2}+W^{2}+AB-C(A+B),~~t=ABC+CW^{2}-AV^{2}.  \label{rst}
\end{equation}%
Reading off the entries for the block in the lower right corner of $%
M_{2}^{2} $ as $A=c_{3}m_{3}^{2}\mu ^{4}/\kappa $, $B=-c_{2}m_{2}^{2}$, $%
C=c_{3}m_{3}^{2}$, $W=c_{\mu }\mu ^{2}$, $V=$ $c_{\nu }\nu ^{2}$, we find
that the constant term in the characteristic equation is zero, i.e. $t=0$.
Hence at least one eigenvalue becomes zero. The remaining equation is simply
quadratic with solutions%
\begin{equation}
\lambda _{\pm }=\frac{c_{3}m_{3}^{2}\mu ^{4}}{2\kappa }-\frac{%
c_{2}m_{2}^{2}+c_{3}m_{3}^{2}}{2}\pm \frac{1}{2\kappa }\sqrt{m_{3}^{4}(\mu
^{4}-\mu _{e}^{4})^{2}+4c_{\nu }\nu ^{2}\kappa ^{3/2}(\mu ^{4}-\mu _{e}^{4})}%
.~~  \label{lpm}
\end{equation}

We introduced here the quantity $\mu _{\text{e}}^{\pm }=[\kappa (\kappa
-m_{3}^{4}+\nu ^{4}\pm 2c_{\nu }\nu ^{2}\sqrt{\kappa })]^{1/4}/m_{3}$, that
signifies the value for $\mu $ at which the eigenvalues $\lambda _{+}$ and $%
\lambda _{-}$ coincide, which is referred to as the exceptional point. For
the block in the top left corner we identify $A=3c_{3}m_{3}^{2}\mu
^{4}/\kappa +2c_{1}m_{1}^{2}$, $B=-c_{2}m_{2}^{2}$, $C=c_{3}m_{3}^{2}$, $%
W=-c_{\mu }\mu ^{2}$ and $V=$ $-c_{\nu }\nu ^{2}$. The linear term becomes $%
t=-2(c_{3}m_{3}^{2}\mu ^{4}+c_{1}m_{1}^{2}\nu
^{4}+c_{1}c_{2}c_{3}m_{1}^{2}m_{2}^{2}m_{3}^{2})$, which is exactly twice
the value of $t$ obtained previously for the vacuum $\Phi _{1}^{0}$. For $%
t\neq 0$ we define with (\ref{rst}) the quantities%
\begin{equation}
\rho =\sqrt{-\frac{p^{3}}{27}},~\cos \theta =-\frac{q}{2\rho },~p=\frac{%
3s-r^{2}}{3},~q=\frac{2r^{3}}{27}-\frac{rs}{3}+t,~\Delta =\left( \frac{p}{3}%
\right) ^{3}+\left( \frac{q}{2}\right) ^{2}.
\end{equation}%
Then, provided that $p<0$ and $\Delta \leq 0$, the remaining three
eigenvalues are real and according to Cardano's formula of the form 
\begin{equation}
\lambda _{i}=2\rho ^{1/3}\cos \left[ \frac{\theta }{3}+\frac{2\pi }{3}(i-1)%
\right] ,\qquad i=1,2,3.
\end{equation}%
Similarly as for the vacuum $\Phi _{1}^{0}$ the values of $c_{\mu }$ and $%
c_{\nu }$ are not relevant for the computation of the eigenvalues.
Naturally, for these eigenvalues to be interpretable as squared masses to
tree order they need to be non-negative. There are indeed some regions in
the parameter space for which this holds, taking for instance $%
c_{1}=c_{3}=-c_{2}=1$, $m_{1}=1$, $m_{2}=1/2$, $m_{3}=1/5$, $\mu =2$ and $%
\nu =1/2$ we compute the six non-negative eigenvalues $(\lambda _{1},\lambda
_{3},\lambda _{2},\lambda _{+},\lambda
_{-},0)=(38.1493,0.5683,0.0639,10.6534,1.7471,0)$. However, as seen in
figure \ref{Fig2} these physical regions are quite isolated in the parameter
space.

\FIGURE{ \epsfig{file=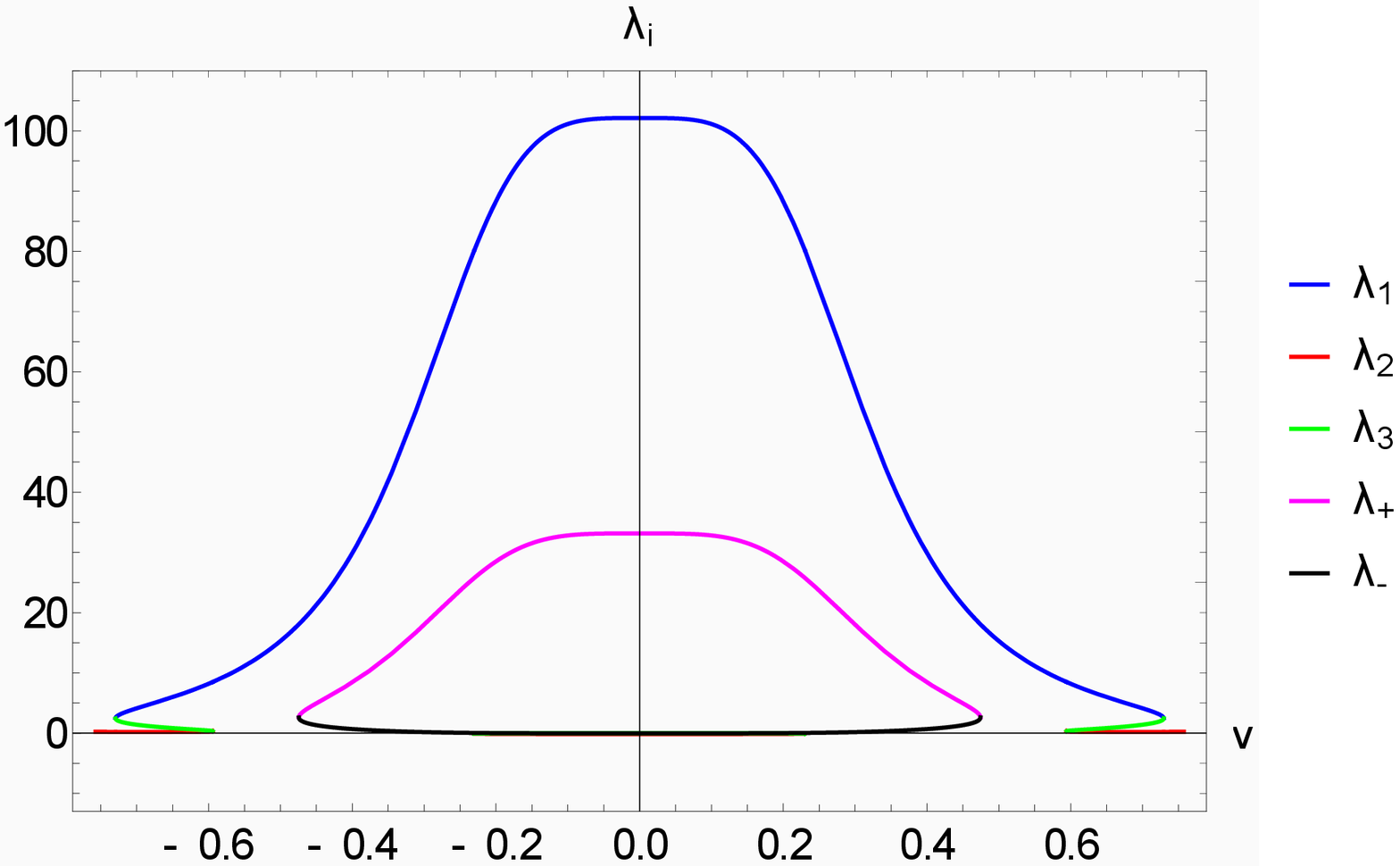,width=7.2cm} \epsfig{file=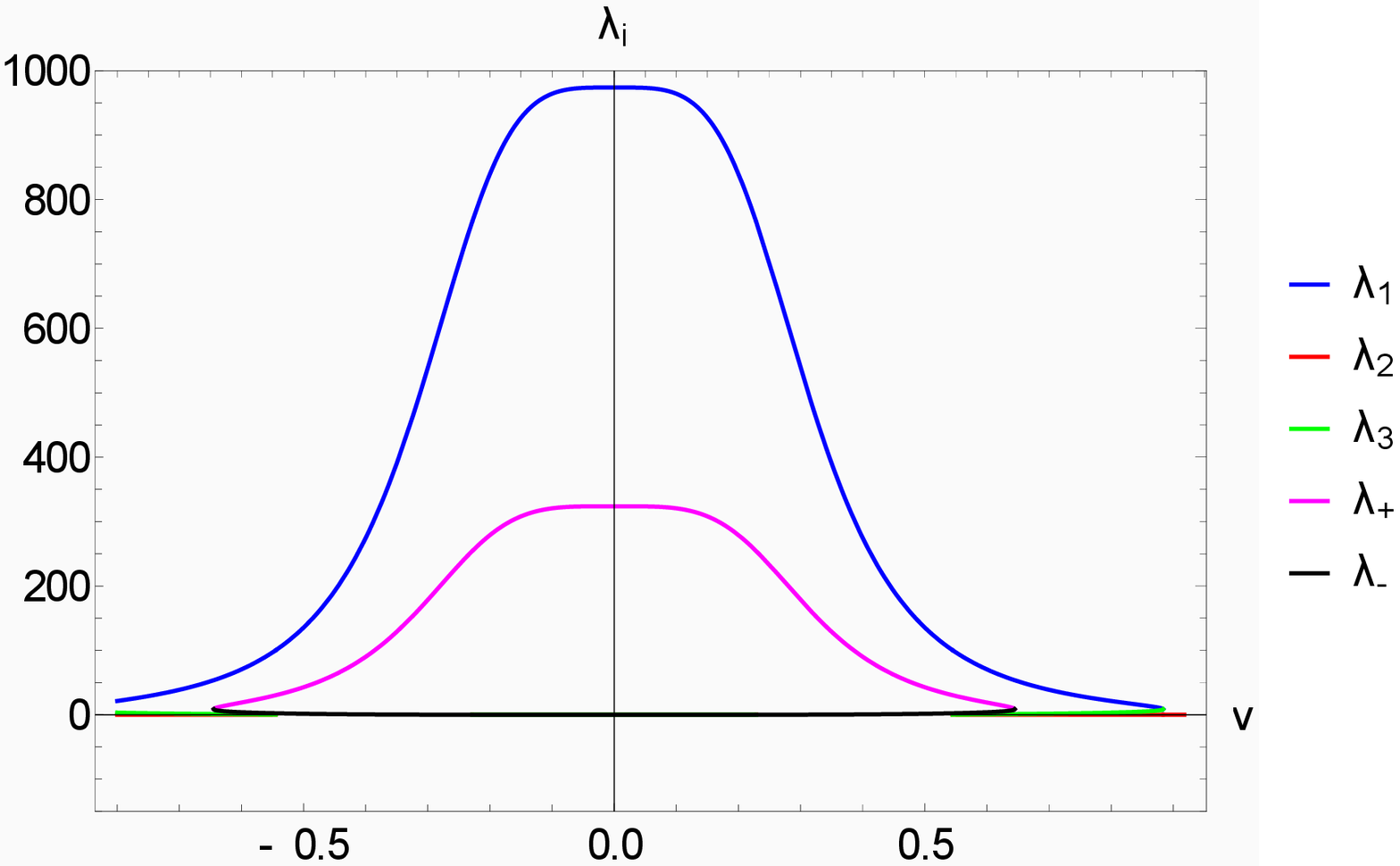,width=7.2cm}
        \caption{Nonvanishing eigenvalues $\lambda_i$ of  $M_{2}^{2}$ as a functions
of $\nu $ for $c_{1}=c_{2}=c_{3}=1$, $m_{1}=1$, $m_{2}=1/2$ and $m_{3}=1/5$. In the left panel we choose $\mu =1.7$ observing that
there is no physical region for which all eigenvalues are non-negative. In the right panel we choose $\mu =3$ and have two physical regions for $\nu \in
(-0.64468,-0.54490)$ and $\nu \in (0.54490,0.64468)$.}
        \label{Fig2}}

For the choice $c_{1}=-c_{3}=\pm 1$ we may also find a value for $\nu =\nu _{%
\text{sing}}^{\pm }=\pm \sqrt{m_{2}m_{3}}$, for which $\kappa \rightarrow 0$
leading to singularities in the eigenvalues. Figure \ref{Fig3} depicts such
a situation.

\FIGURE{ \epsfig{file=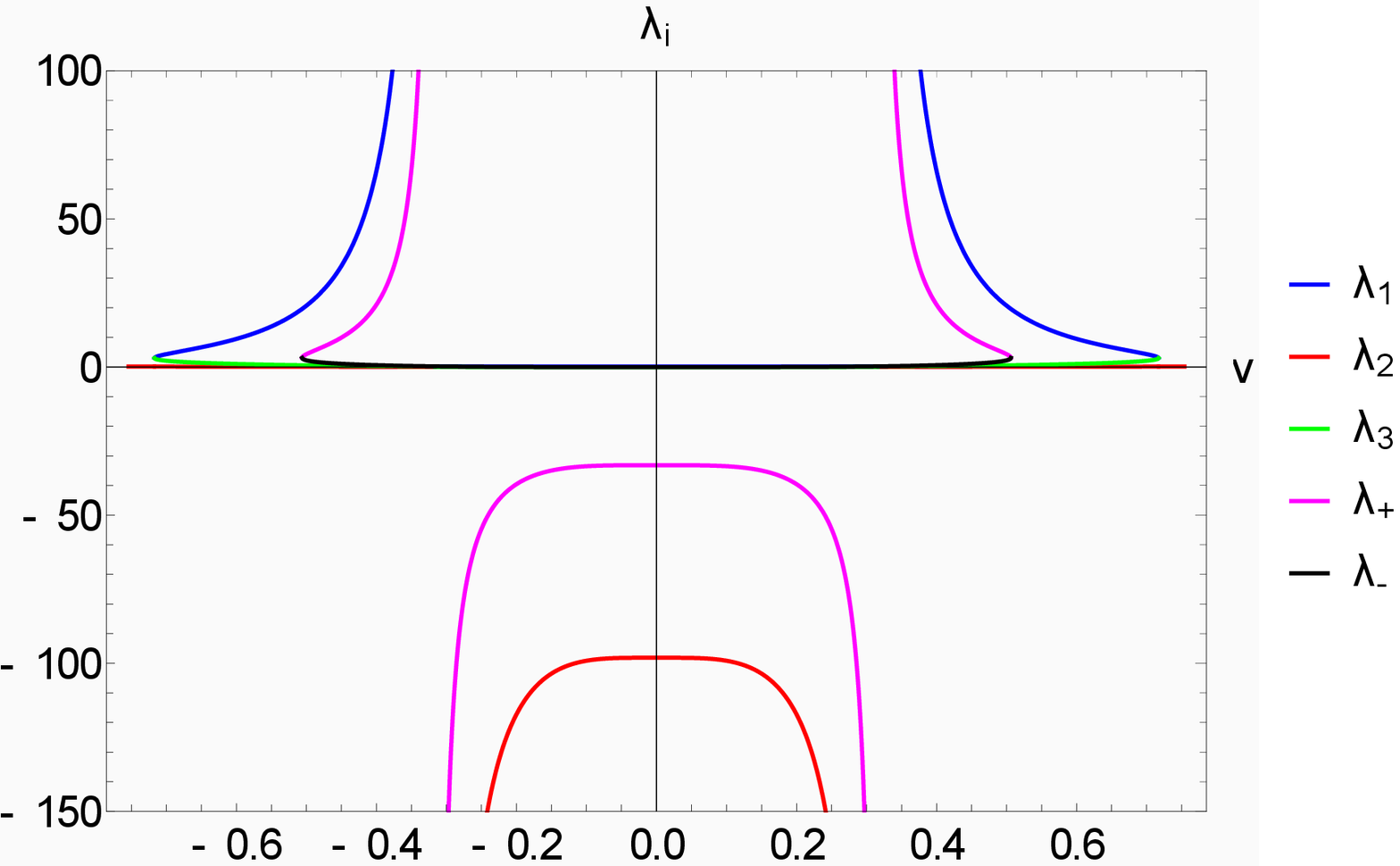,width=11.0cm} 
\caption{Nonvanishing eigenvalues $\lambda_i$ of  $M_{2}^{2}$ as a function
of $\nu $ for $c_{1}=-c_{2}=c_{3}=1$, $m_{1}=1$, $m_{2}=1/2$,$m_{3}=1/5$ and $\mu =1.7$. 
Singularities occur at $\nu=\nu _{\text{sing}}^{\pm } \approx \pm  0.31623$. The regimes $\nu \in
(-0.50608,\nu _{\text{sing}}^{-})$, $\nu \in (\nu _{\text{sing}}^{+},0.50608)
$ are physical.}
        \label{Fig3}}

As for the case with $U(1)$-invariant vacuum, for some specific choices of $%
\mu $ we can apparently generate an additional massless particle. Since the
linear term of the characteristic equation for the upper right corner is
simply twice the one of the previous section, this scenario occurs for $\mu
=\mu _{s}$. However, as we pointed out above for this value of $\mu $ the
two vacua $\Phi _{1}^{0}$ and $\Phi _{2}^{0}$ coincide, so that the
discussion of the previous section applies. In addition, as the two blocks
are different in this case there is a second choice $\bar{\mu}%
_{s}^{4}=\kappa ^{2}/(m_{3}^{4}-\nu ^{4})$ for which $\lambda _{-}=0$ and
the non-zero eigenvalue coalesces with the zero eigenvalue at the
zero-exceptional point. Hence, in this case it appears that besides the
Goldstone boson there is a second massless, non-Goldstone, particle present
in the model. We shall see below that this is actually not the case.

Choosing instead the vacuum $\Phi _{3}^{0}$, the resulting mass matrix $%
M_{3}^{2}$ is similar to $M_{2}^{2}$ with the block in the top left corner
and lower right corner exchanged accompanied by the transformation $c_{\nu
/\mu }\rightarrow -c_{\nu /\mu }$, hence the previous discussion applied in
this case.

Expanding instead around the vacuum $\Phi _{4}^{0}$ the resulting mass
matrix reads%
\begin{equation}
M_{4}^{2}=\left( 
\begin{array}{cccccc}
\frac{c_{3}m_{3}^{2}\mu ^{4}}{\kappa }+(\varphi _{1}^{0})^{2} & -\frac{g}{2}%
c_{\mu }\mu ^{2} & 0 & \frac{g}{2}\varphi _{1}^{0}\chi _{1}^{0} & 0 & 0 \\ 
c_{\mu }\mu ^{2} & -c_{2}m_{2}^{2} & c_{\nu }\nu ^{2} & 0 & 0 & 0 \\ 
0 & -c_{\nu }\nu ^{2} & -c_{3}m_{3}^{2} & 0 & 0 & 0 \\ 
\frac{g}{2}\varphi _{1}^{0}\chi _{1}^{0} & 0 & 0 & 2c_{1}m_{1}^{2}+\frac{%
3c_{3}m_{3}^{2}\mu ^{4}}{\kappa }-\frac{g}{2}(\varphi _{1}^{0})^{2} & c_{\mu
}\mu ^{2} & 0 \\ 
0 & 0 & 0 & -c_{\mu }\mu ^{2} & -c_{2}m_{2}^{2} & -c_{\nu }\nu ^{2} \\ 
0 & 0 & 0 & 0 & c_{\nu }\nu ^{2} & -c_{3}m_{3}^{2}%
\end{array}%
\right) .
\end{equation}%
Computing the sixth order characteristic polynomial for $M_{4}^{2}$ we find
that the dependence on the free field $\varphi _{1}^{0}$ drops out entirely.
We also note that the linear term always vanishes and that therefore a
Goldstone boson is present for this vacuum. We will not present here a more
detailed discussion as the qualitative behaviour of the model is similar to
the one discussed in detail in the previous section. The model posses
various well defined physical regions. For instance, for $c_{1}=-1$, $%
c_{2}=c_{3}=c_{\mu }=c_{\nu }=1$, $m_{1}=2$, $m_{2}=1/2$, $m_{3}=1/10$, $\mu
=3/2$ and $\nu =0.28$ we find the eigenvalues $%
(0,0.0130,0.2731,0.7294,4.8655,9.0186)$ for $M_{4}^{2}$. Let us now see how
to explain the reality of the mass spectrum.

\subsection{From quantum mechanical to field theoretical $\mathcal{P}$%
-operators}

We consider now the lower right block of the squared mass matrix in (\ref%
{mv2}) and construct a $\mathcal{P}$-operator in a manner as describes in
section 3.2, i.e. taking the mass matrix as a starting point. Subsequently
we verify whether the operator constructed in the manner is a parity
operator that can be used in the $\mathcal{CPT}$-symmetry transformations
that leave the quantum field theoretical actions invariant. Including the
remaining part of the squared mass matrix is straightforward.

We consider the version of $M_{2}^{2}$ resulting from the action before
carrying out the similarity transformation, with the lower right block in (%
\ref{mv2}) given as 
\begin{equation}
\mathcal{M}=\left( 
\begin{array}{ccc}
\frac{c_{3}m_{3}^{2}\mu ^{4}}{\kappa } & ic_{\mu }\mu ^{2} & 0 \\ 
ic_{\mu }\mu ^{2} & -c_{2}m_{2}^{2} & -ic_{\nu }\nu ^{2} \\ 
0 & -ic_{\nu }\nu ^{2} & -c_{3}m_{3}^{2}%
\end{array}%
\right) .
\end{equation}%
The standard argument that explains the reality of the spectrum for this
non-Hermitian matrix is simply stated: Iff there exists an antilinear
operator $\mathcal{PT}$, satisfying%
\begin{equation}
\left[ \mathcal{M},\mathcal{PT}\right] =0,\qquad \text{and\qquad }\mathcal{PT%
}v_{n}=v_{n}  \label{anti}
\end{equation}%
with $v_{n}$ denoting the eigenvectors of $\mathcal{M}$, the eigenvalues $%
\lambda _{n}$ of $\mathcal{M}$ are real. When in (\ref{anti}) only the first
relation holds and $\mathcal{PT}v_{n}\neq v_{n}$, the $\mathcal{PT}$%
-symmetry is spontaneously broken and some of the eigenvalues emerge in
complex conjugate pairs.

To check this statement for our concrete matrix and in particular to
construct an explicit expression for the $\mathcal{P}$-operator we compute
first the normalised left and right eigenvectors for this non-Hermitian
matrix as defined in (\ref{leftright})%
\begin{equation}
v_{j}=(-1)^{\delta _{-,j}}u_{j}^{\ast }=\frac{1}{N_{j}}\{-\lambda
_{j}\Lambda _{j}-\kappa ,-i\Lambda _{j}^{3}c_{\mu }\mu ^{2},-c_{\mu }c_{\nu
}\mu ^{2}\nu ^{2}\},~~~j=0,\pm ,
\end{equation}%
with normalisation constants%
\begin{eqnarray}
N_{\pm }^{2} &=&(\kappa +\lambda _{\pm }\Lambda _{\pm })\lambda _{\pm
}\left( \lambda _{+}-\lambda _{-}\right) , \\
N_{0}^{2} &=&\kappa \lambda _{-}\lambda _{+},
\end{eqnarray}%
where we abbreviated $\Lambda _{j}:=\lambda
_{j}+c_{2}m_{2}^{2}+c_{3}m_{3}^{2}$ and $\Lambda _{j}^{k}:=\lambda
_{j}+c_{k}m_{k}^{2}$. We confirm that the set of vectors $\{v_{j},u_{j}\}$
with $j=0,\pm $ form indeed a biorthonormal basis by verifying (\ref{uv}).

Next we use relation (\ref{PPT}) to compute the $\mathcal{P}$-operator%
\begin{equation}
\mathcal{P=}\sum\limits_{j=0,\pm }\frac{s_{j}}{N_{j}^{2}}\left( 
\begin{array}{ccc}
\left( \Lambda _{j}^{2}\Lambda _{j}^{3}+\nu ^{4}\right) ^{2} & i\mu
^{2}\Lambda _{j}^{3}\left( \Lambda _{j}^{2}\Lambda _{j}^{3}+\nu ^{4}\right)
& \mu ^{2}\nu ^{2}\left( \Lambda _{j}^{2}\Lambda _{j}^{3}+\nu ^{4}\right) \\ 
-i\mu ^{2}\Lambda _{j}^{3}\left( \Lambda _{j}^{2}\Lambda _{j}^{3}+\nu
^{4}\right) & \mu ^{4}\left( \Lambda _{j}^{3}\right) ^{2} & -i\nu ^{2}\mu
^{4}\Lambda _{j}^{3} \\ 
\mu ^{2}\nu ^{2}\left( \Lambda _{j}^{2}\Lambda _{j}^{3}+\nu ^{4}\right) & 
i\nu ^{2}\mu ^{4}\Lambda _{j}^{3} & \mu ^{4}\nu ^{4}%
\end{array}%
\right)
\end{equation}%
Given all possibilities for the signatures $s_{n}$, we have found eight
different $\mathcal{P}$-operators. All of them satisfy the two relations in (%
\ref{PTM}). However, two signatures are very special as for them the
expressions simplify considerably 
\begin{equation}
\mathcal{P}(s_{0}=\pm 1,s_{-}=\mp 1,s_{+}=\pm 1)=\left( 
\begin{array}{ccc}
\pm 1 & 0 & 0 \\ 
0 & \mp 1 & 0 \\ 
0 & 0 & \pm 1%
\end{array}%
\right) .
\end{equation}%
Moreover, in this case the $\mathcal{P}$-operators are indeed the operators
involved in the $\mathcal{CPT}_{1/2}$-symmetry transformation that is
respected by the entire action. Notice that at the exceptional point, $%
\lambda _{-}=\lambda _{+}$, the normalisation factors $N_{\pm }$ becomes
zero so that the eigenvector $v_{\pm }$ and $u_{\pm }$ are no longer
defined. Passing this point corresponds to breaking the $\mathcal{PT}$%
-symmetry spontaneously and the second relation in (\ref{anti}) no longer
holds.

Next we calculate the operator $C$ as defined in equation (\ref{C}) in two
alternative ways to

\begin{equation}
C=\sum\limits_{j=0,\pm }\frac{(-1)^{\delta _{-,j}}s_{j}}{N_{j}^{2}}\left( 
\begin{array}{ccc}
\left( \Lambda _{j}^{2}\Lambda _{j}^{3}+\nu ^{4}\right) ^{2} & i\mu
^{2}\Lambda _{j}^{3}\left( \Lambda _{j}^{2}\Lambda _{j}^{3}+\nu ^{4}\right)
& \mu ^{2}\nu ^{2}\left( \Lambda _{j}^{2}\Lambda _{j}^{3}+\nu ^{4}\right) \\ 
i\mu ^{2}\Lambda _{j}^{3}\left( \Lambda _{j}^{2}\Lambda _{j}^{3}+\nu
^{4}\right) & -\mu ^{4}\left( \Lambda _{j}^{3}\right) ^{2} & i\nu ^{2}\mu
^{4}\Lambda _{j}^{3} \\ 
\mu ^{2}\nu ^{2}\left( \Lambda _{j}^{2}\Lambda _{j}^{3}+\nu ^{4}\right) & 
i\nu ^{2}\mu ^{4}\Lambda _{j}^{3} & \mu ^{4}\nu ^{4}%
\end{array}%
\right) .
\end{equation}%
We verify that $C$ does indeed satisfy all the relations in (\ref{Calg}).
The Dyson operator is identified as $\eta =U^{-1}$ with $%
U=(v_{0},v_{+},v_{-})$ and the metric operator as $\rho =\eta ^{\dagger
}\eta $. Since $\det U=i\lambda _{-}\lambda _{+}(\lambda _{-}-\lambda
_{+})\mu ^{4}\nu ^{2}/$ $N_{0}N_{-}N_{+}$ both operators exist in the $%
\mathcal{PT}$-symmetric regime. The fact that the $C$-operator is not unique 
\cite{bender2009nonunique} is a well known fact, similarly as for the metric
operator.

\subsection{The Goldstone boson in the $\mathcal{PT}$-symmetric regime}

Let us now compute the explicit expression for the Goldstone boson. As we
have seen in section 3.5, the Goldstone boson emerges from the lower right
block so that it suffices to consider that part of the squared mass matrix.
Denoting the quantities related to the lower right block by a subscript $r$
and the upper left block by $\ell $, we decompose the Lagrangian into $%
\mathcal{L}_{3}=\mathcal{L}_{3,\ell }+\mathcal{L}_{3,r}$ and define the
quantities 
\begin{equation}
\hat{\Phi}_{r}:=(\hat{\chi}_{1},\hat{\varphi}_{2},\hat{\chi}_{3}),\qquad
(M_{2}^{2})_{r}v_{i}=\lambda _{i}v_{i},\qquad U:=(v_{0},v_{+},v_{-}),\quad
i=0,\pm .  \label{FMU}
\end{equation}%
Similarly for $\mathcal{L}_{3,\ell }$, which we, however, do not analyse
here as it does not contain a Goldstone boson. Thus, as long as the spectrum
of $M_{2}^{2}$ is not degenerate, and hence all the eigenvectors $v_{i}$ are
linearly independent, the matrix $U$ diagonalizes the lower right block of
the squared mass matrix $U^{-1}(M_{2}^{2})_{r}U=D$ with $\limfunc{diag}%
D=(\lambda _{0},\lambda _{+},\lambda _{-})=(m_{0}^{2},m_{+}^{2},m_{-}^{2})$.
As argued in general in (\ref{psim})-(\ref{psi}), we may therefore defined
the fields $\psi _{k}$, $k=0,\pm $, with masses $m_{i}$ by re-writing the
mass term%
\begin{equation}
\hat{\Phi}_{r}^{T}(M_{2}^{2})_{r}\hat{\Phi}_{r}=\sum\nolimits_{k=0,\pm
}m_{k}^{2}\psi _{k}^{2}=\sum\nolimits_{k=0,\pm }m_{k}^{2}(\hat{\Phi}%
_{r}^{T}IU)_{k}(U^{-1}\Phi _{r})_{k}.
\end{equation}%
Hence, the Goldstone field corresponding to $\psi _{0}$ is expressible as%
\begin{equation}
\psi _{\text{Gb}}:=\sqrt{\left( \hat{\Phi}_{r}^{T}IU\right) _{0}(U^{-1}\hat{%
\Phi}_{r})_{0}}.  \label{FB}
\end{equation}%
The unnormalised right eigenvectors for $M_{2}^{2}$ are computed to%
\begin{equation}
v_{i}=\{-\lambda _{i}\Lambda _{i}-\kappa ,\Lambda _{i}^{3}c_{\mu }\mu
^{2},c_{\mu }c_{\nu }\mu ^{2}\nu ^{2}\},~~~i=0,\pm ,
\end{equation}%
so that the explicit form of the Goldstone boson field in the original
fields becomes%
\begin{equation}
\psi _{\text{Gb}}:=\frac{1}{\sqrt{N}}\left( -\kappa \hat{\chi}%
_{1}-c_{3}c_{\mu }m_{3}^{2}\mu ^{2}\hat{\varphi}_{2}+c_{\mu }c_{\nu }\mu
^{2}\nu ^{2}\hat{\chi}_{3}\right) ,  \label{FG}
\end{equation}%
with%
\begin{equation}
N=m_{3}^{4}(m_{2}^{4}-\mu ^{4})+(2c_{2}c_{3}m_{2}^{2}m_{3}^{2}+\mu ^{4})\nu
^{4}+\nu ^{8}=\kappa ^{2}\left( 1-\frac{\mu ^{2}}{\bar{\mu}_{s}^{2}}\right) ,
\end{equation}%
where $\bar{\mu}_{s}$ is defined as above being the special value of $\mu $
for which $\lambda _{-}=0$, that is the zero-exceptional point. Computing
the determinant of $U$ to $\det U=c_{\nu }\lambda _{-}\lambda _{p}(\lambda
_{-}-\lambda _{p})\nu ^{2}\mu ^{4}$, the origin of this singularity is
clear, as $U$ is not invertible for vanishing for $\lambda _{-}=0$ and at
the standard exceptional points when $\lambda _{-}=\lambda _{p}$. The former
scenario occurs for $\mu =$ $\bar{\mu}_{s}$ and the latter for $\mu _{\text{e%
}}^{\pm }=[\kappa (\kappa -m_{3}^{4}+\nu ^{4}\pm 2c_{\nu }\nu ^{2}\sqrt{%
\kappa })]^{1/4}/m_{3}$. So that in these circumstances the Goldstone boson
of the form (\ref{FG}) does not exist. We discuss these two scenarios
separately in the next two sections. However, for $\mu =\mu _{s}$, that is
the value for which the other sector develops a massless particle, all terms
in $\psi _{\text{Gb}}$ are regular. This means at this point we have two
massless particles in the model. One is tempted to interpret one as a
genuine Goldstone boson and the other as simple massless particle. However,
recalling that at $\mu =\mu _{s}$ one is actually expanding around the $U(1)$%
-symmetry preserving vacuum the emergence of none of them can be attributed
to a global symmetry breaking and the discussion in section 3.3 applies.

\subsection{The Goldstone boson at the exceptional point}

As pointed out in the previous section, at the exceptional point when $%
\lambda _{-}=\lambda _{p}=:\lambda _{e}$ the matrix $U$ is no longer
invertible so that $\psi _{\text{Gb}}$ in (\ref{FB}) becomes ill-defined.
However, when $\mu =$ $\mu _{\text{e}}^{+}=\mu _{\text{e}}$ we may transform
the lower right block of $M_{2}^{2}$ into Jordan normal form as%
\begin{equation}
T^{-1}\left[ M_{2}^{2}(\mu =\mu _{\text{e}})\right] _{r}T=\left( 
\begin{array}{lll}
0 & 0 & 0 \\ 
0 & \lambda _{\text{e}} & a \\ 
0 & 0 & \lambda _{\text{e}}%
\end{array}%
\right) =J,  \label{TJ}
\end{equation}%
for some as yet unspecified constant $a\in \mathbb{R}$. For simplicity we
select here the upper sign of the two possibilities $\mu _{\text{e}}^{\pm }$%
. We can then express the transformed action expanded around the vacuum $%
\Phi _{2}^{0}$ and formulate the Goldstone boson in terms of the original
fields 
\begin{eqnarray}
\hat{{\mathcal{I}}}_{3} &=&-\frac{1}{2}\int d^{4}x\left[ \hat{\Phi}%
^{T}I(\square +M_{2}^{2})\hat{\Phi}+\mathcal{L}_{\text{int}}(\hat{\Phi})+%
\mathcal{L}_{3,\ell }\right] , \\
&=&-\frac{1}{2}\int d^{4}x\left[ \hat{\Phi}^{T}IT(\square +J)T^{-1}\hat{\Phi}%
+\mathcal{L}_{\text{int}}(\hat{\Phi})+\mathcal{L}_{3,\ell }\right] , \\
&=&-\frac{1}{2}\int d^{4}x\left[ \sum\limits_{i=1}^{3}\psi _{i}\square \psi
_{i}+\lambda _{e}(\psi _{2}^{2}+\psi _{3}^{2})+a\psi _{2}^{L}\psi _{3}^{R}+%
\mathcal{L}_{\text{int}}(\psi _{i})+\mathcal{L}_{3,\ell }\right] .
\end{eqnarray}%
We have introduced here the fields 
\begin{equation}
\psi _{i}:=\sqrt{\psi _{i}^{L}\psi _{i}^{R}},\qquad \psi _{i}^{L}:=(\hat{\Phi%
}_{r}^{T}IT)_{i},\qquad \psi _{i}^{R}:=(T^{-1}\hat{\Phi}_{r})_{i},
\label{goldex}
\end{equation}%
with the Goldstone boson at the exceptional point being identified as $\psi
_{\text{Gb}}^{\text{e}}:=\psi _{1}$. Notice that when $T^{T}IT=\mathbb{I}$,
the field coincide, i.e. we have $\psi _{i}^{L}=\psi _{i}^{R}=\psi _{i}$.
Let us now determine the matrix $T$ and demonstrate that it is well-defined.
We take $\mu =\mu _{\text{e}}$ so that the nonzero eigenvalue for $%
M_{2}^{2}(\mu _{\text{e}})$ becomes 
\begin{equation}
\lambda _{\text{e}}=\frac{\nu ^{4}-m_{3}^{4}+c_{\nu }\nu ^{2}\sqrt{\kappa }}{%
c_{3}m_{3}^{2}}.
\end{equation}%
Using the null vector of $M_{2}^{2}(\mu _{\text{e}})$ and the eigenvector
corresponding to the eigenvalue $\lambda _{\text{e}}$ in the first and
second column of $T$, respectively, we solve equation (\ref{TJ}) for $T$ as%
\begin{equation}
T=\left( 
\begin{array}{ccc}
-\kappa c_{3}m_{3}^{2} & -c_{3}m_{3}^{2}\mu _{\text{e}}^{2} & t \\ 
m_{3}^{4}\mu _{\text{e}}^{2} & \kappa +c_{\nu }\nu ^{2}\sqrt{\kappa } & s \\ 
c_{3}c_{\nu }\nu ^{2}m_{3}^{2}\mu _{\text{e}}^{2} & c_{3}m_{3}^{2}\sqrt{%
\kappa } & \frac{s-\sqrt{\kappa }}{c_{3}m_{3}^{2}+\lambda _{\text{e}}}\nu
^{2}%
\end{array}%
\right) ,
\end{equation}%
with abbreviations $t:=(1-m_{3}^{4}-\nu ^{4})\mu _{\text{e}}^{2}/(\lambda _{%
\text{e}}\sqrt{\kappa })$, $s:=t\left( \lambda _{\text{e}}/\mu _{\text{e}%
}^{2}-c_{3}m_{3}^{2}\mu _{\text{e}}^{2}/\kappa \right) -\nu ^{2}$ and $a$ as
defined in (\ref{TJ}) taken to $a=\nu ^{2}/m_{3}^{6}$. We compute $\det
T=\kappa m_{3}^{4}\lambda _{\text{e}}^{2}\,$. We have imposed here $\psi
_{1}^{L}=\psi _{1}^{R}=\psi _{1}$. Using these expression we obtain from (%
\ref{goldex}) the Goldstone boson at the exceptional point as%
\begin{equation}
\psi _{\text{Gb}}^{\text{e}}=\frac{1}{\kappa c_{3}m_{3}^{2}\lambda _{\text{e}%
}^{2}}\left( -\kappa \hat{\chi}_{1}-m_{3}\mu _{\text{e}}^{2}\hat{\varphi}%
_{2}+\nu ^{2}\mu _{\text{e}}^{2}\hat{\chi}_{3}\right) .
\end{equation}%
Thus at the exceptional point the Goldstone boson $\psi _{\text{Gb}}^{\text{e%
}}$ is well-defined unless $\lambda _{\text{e}}=0$, $\kappa =0$ or $m_{3}=0$%
, as in these cases the matrix $T$ is not invertible.

\subsection{The Goldstone boson at the zero-exceptional point}

Another interesting point at which the general expression for the Goldstone
boson in (\ref{FB}) is not valid occurs for $\mu =$ $\bar{\mu}_{s}$, that is
when $\lambda _{-}=0$, i.e. at the zero-exceptional point. In this case we
may transform the lower right block of $M_{2}^{2}$ into the form%
\begin{equation}
S^{-1}\left[ M_{2}^{2}(\mu =\bar{\mu}_{s})\right] _{r}S=\left( 
\begin{array}{lll}
0 & 0 & b \\ 
0 & \lambda _{\text{s}} & 0 \\ 
0 & 0 & 0%
\end{array}%
\right) =K,  \label{SK}
\end{equation}%
for some as yet unspecified constant $b\in \mathbb{R}$. As before we can
then express the transformed action expanded around the vacuum $\Phi
_{2}^{0} $ and formulate the Goldstone boson in terms of the original fields 
\begin{eqnarray}
\hat{{\mathcal{I}}}_{3} &=&-\frac{1}{2}\int d^{4}x\left[ \hat{\Phi}%
^{T}I(\square +M_{2}^{2})\hat{\Phi}+\mathcal{L}_{\text{int}}(\hat{\Phi})+%
\mathcal{L}_{3,\ell }\right] , \\
&=&-\frac{1}{2}\int d^{4}x\left[ \hat{\Phi}^{T}IS(\square +K)S^{-1}\hat{\Phi}%
+\mathcal{L}_{\text{int}}(\hat{\Phi})+\mathcal{L}_{3,\ell }\right] , \\
&=&-\frac{1}{2}\int d^{4}x\left[ \sum\limits_{i=1}^{3}\psi _{i}\square \psi
_{i}+\lambda _{s}\psi _{2}^{2}+b\psi _{1}^{L}\psi _{3}^{R}+\mathcal{L}_{%
\text{int}}(\psi _{i})+\mathcal{L}_{3,\ell }\right] ,
\end{eqnarray}%
where we introduced 
\begin{equation}
\psi _{i}:=\sqrt{\psi _{i}^{L}\psi _{i}^{R}},\qquad \psi _{i}^{L}:=(\hat{\Phi%
}_{r}^{T}IS)_{i},\qquad \psi _{i}^{R}:=(S^{-1}\hat{\Phi}_{r})_{i}.
\end{equation}%
Taking $\mu =\bar{\mu}_{s}$, the only nonzero eigenvalue for $M_{2}^{2}(\bar{%
\mu}_{s})$ becomes 
\begin{equation}
\lambda _{\text{z}}=\frac{\left( c_{2}m_{2}^{2}+2c_{3}m_{3}^{2}\right) \nu
^{4}-c_{3}m_{3}^{6}}{m_{3}^{4}-\nu ^{4}}.
\end{equation}%
Using the null vector of $M_{2}^{2}(\bar{\mu}_{s})$ and the eigenvector
corresponding to the eigenvalue $\lambda _{\text{e}}$ in the first and
second column of $S$, respectively, we solve equation (\ref{SK}) for $S$ to%
\begin{equation}
S=\left( 
\begin{array}{ccc}
-\sqrt{m_{3}^{4}-\nu ^{4}} & -\nu ^{2}\kappa & 0 \\ 
c_{3}m_{3}^{2} & \left( c_{2}m_{2}^{2}+c_{3}m_{3}^{2}\right) \nu ^{2}\sqrt{%
m_{3}^{4}-\nu ^{4}} & \frac{b}{\kappa }(\nu ^{4}-m_{3}^{4}) \\ 
\nu ^{2} & (m_{3}^{4}-\nu ^{4})^{3/2} & -\frac{b}{\kappa }\left(
c_{2}m_{2}^{2}+c_{3}m_{3}^{2}\right) \nu ^{2}%
\end{array}%
\right) .
\end{equation}%
We compute $\det S=-b\lambda _{\text{z}}^{2}\,(m_{3}^{4}-\nu
^{4})^{2}/\kappa $. The massive field $\psi _{2}$ can be identified easily
for any value of $b$ as%
\begin{equation}
\psi _{2}=\frac{1}{N_{2}}\psi _{2}^{L}
\end{equation}%
when noting that 
\begin{equation}
\psi _{2}^{L}=N_{2}^{2}\psi _{2}^{R}=-\kappa \nu ^{2}\hat{\chi}_{1}-\left(
c_{2}m_{2}^{2}+c_{3}m_{3}^{2}\right) \nu ^{2}\sqrt{m_{3}^{4}-\nu ^{4}}\hat{%
\varphi}_{2}+(m_{3}^{4}-\nu ^{4})^{3/2}\hat{\chi}_{3},
\end{equation}%
with $N_{2}=(m_{3}^{4}-\nu ^{4})\lambda _{\text{z}}$. However, we can not
identify the Goldstone boson simply as $\psi _{1}$, since we can no longer
achieve $\psi _{1}^{L}\propto \psi _{1}^{R}\propto \psi _{1}$. Given the
eigenvalue spectrum we have now two massless particles that interact with
each other and it is impossible to distinguish the Goldstone boson from the
massless particle. However, we can identify a combination of the two fields
as a massless particle%
\begin{eqnarray}
\psi _{\text{Gb}}^{\text{z}} &=&\psi _{1}^{L}+\alpha \psi _{3}^{L}=\psi
_{1}^{R}+\alpha \psi _{3}^{R} \\
&=&-\sqrt{m_{3}^{4}-\nu ^{4}}\hat{\chi}_{1}+\frac{\left( m_{3}^{4}-\nu
^{4}\right) ^{2}+\nu ^{4}(1-\kappa )-m_{3}^{4}}{(m_{3}^{4}-\nu ^{4})\lambda
_{\text{z}}}\hat{\varphi}_{2}+\nu ^{2}\left[ 1+\frac{%
c_{2}m_{2}^{2}+c_{3}m_{3}^{2}}{(m_{3}^{4}-\nu ^{4})\lambda _{\text{z}}}%
\right] \hat{\chi}_{3},  \notag
\end{eqnarray}%
for $b=$ $-\bar{\mu}_{s}^{4}/(\alpha \kappa \lambda _{\text{z}})$ and $%
\alpha ^{2}=1+(\bar{\mu}_{s}^{4}-m_{2}^{4}-m_{3}^{4}+2\nu ^{4})/[\lambda _{%
\text{z}}^{2}(m_{3}^{4}-\nu ^{4})]$. However, we can not avoid that
constituents of the field, that is $\psi _{1}^{L}$ and $\psi _{3}^{R}$,
interact with each other. The peculiar behaviour at the zero-exceptional
point was also discussed by Mannheim \cite{mannheim2018goldstone} in the
context of the $\mathcal{I}_{2}$-model.

\section{Discrete antilinear and broken continuous global symmetry}

Next we study a non-Hermitian $\mathcal{CPT}$-invariant action but with
broken continuous global $U(1)$-symmetry. This is achieved by keeping in the
Lagrangian density functional (\ref{actionan}) only the two complex scalar
fields $\phi _{1}$ and $\phi _{2}$ genuinely complex and taking the field $%
\phi _{3}$ to be real. Hence we consider the Lagrangian density functional 
\begin{eqnarray}
\mathcal{L}_{3}^{\prime } &=&\sum\limits_{i=1}^{2}\left( \partial _{\mu
}\phi _{i}\partial ^{\mu }\phi _{i}^{\ast }+c_{i}m_{i}^{2}\phi _{i}\phi
_{i}^{\ast }\right) +\left( \partial _{\mu }\phi _{3}\partial ^{\mu }\phi
_{3}+c_{3}m_{3}^{2}\phi _{3}^{2}\right)   \label{L3} \\
&&+c_{\mu }\mu ^{2}\left( \phi _{1}^{\ast }\phi _{2}-\phi _{2}^{\ast }\phi
_{1}\right) +c_{\nu }\nu ^{2}\phi _{3}\left( \phi _{2}-\phi _{2}^{\ast
}\right) -\frac{g}{4}(\phi _{1}\phi _{1}^{\ast })^{2}.  \notag
\end{eqnarray}%
Clearly this model is still $\mathcal{CPT}_{1,2\text{ }}$-invariant, but due
to the presence of the real scalar field the continuous global $U(1)$%
-symmetry is broken already at the level of the action. We parameterize $%
\phi _{i}=1/\sqrt{2}(\varphi _{i}+i\chi _{i})$ with $\varphi _{i}$, $\chi
_{i}\in \mathbb{R}$ for $i=1,2$ and $\phi _{3}=\varphi _{3}/\sqrt{2}$.
Defining the vector field $\Phi :=(\varphi _{1},\chi _{2},\varphi _{3},\chi
_{1},\varphi _{2})^{T}$ and the diagonal $5\times 5$-matrix $E$ with $%
\limfunc{diag}E=(1,0,0,1,0)$, we can write $\mathcal{L}_{3}^{\prime }$ with
the real field content in the compact form%
\begin{equation}
\mathcal{L}_{3}^{\prime }=\frac{1}{2}\partial _{\mu }\Phi ^{T}\partial ^{\mu
}\Phi -\frac{1}{2}\Phi ^{T}M^{2}\Phi -\frac{g}{16}\left( \Phi ^{T}E\Phi
\right) ^{2},
\end{equation}%
with complex mass matrix%
\begin{equation}
M^{2}=\left( 
\begin{array}{ccccc}
-c_{1}m_{1}^{2} & -ic_{\mu }\mu ^{2} & 0 & 0 & 0 \\ 
-ic_{\mu }\mu ^{2} & -c_{2}m_{2}^{2} & -ic_{\nu }\nu ^{2} & 0 & 0 \\ 
0 & -ic_{\nu }\nu ^{2} & -c_{3}m_{3}^{2} & 0 & 0 \\ 
0 & 0 & 0 & -c_{1}m_{1}^{2} & ic_{\mu }\mu ^{2} \\ 
0 & 0 & 0 & ic_{\mu }\mu ^{2} & -c_{2}m_{2}^{2}%
\end{array}%
\right) .  \label{M3}
\end{equation}%
As in the previous section, we similarity transform the corresponding action
using the same Dyson map (\ref{eta}), hence obtaining%
\begin{equation}
\hat{{\mathcal{I}}}_{3}^{\prime }=\eta \mathcal{I}_{3}^{\prime }\eta
^{-1}=\int d^{4}x\left[ \frac{1}{2}\partial _{\mu }\Phi ^{T}I\partial ^{\mu
}\Phi -\frac{1}{2}\Phi ^{T}H\Phi -\frac{g}{16}\left( \Phi ^{T}E\Phi \right)
^{2}\right] ,  \label{I3d}
\end{equation}%
with $H$ being identical to $M^{2}$ in (\ref{M3}) with all imaginary units $i
$ removed. The equation of motion resulting from (\ref{I3d}) reads%
\begin{equation}
-\square I\Phi -H\Phi -\frac{g}{4}\left( \Phi ^{T}E\Phi \right) E\Phi =0,
\end{equation}%
from which we identify the mass matrix as $\hat{M}^{2}=IH$ and by solving $%
\delta V=0$ we obtain the five vacua%
\begin{eqnarray}
\Phi _{0}^{(0)} &:&=(0,0,0,0,0)^{T}, \\
\Phi _{0}^{(1\pm )} &:&=\frac{2}{m_{2}}\sqrt{\frac{\kappa }{c_{2}g}}\left(
0,0,0,\pm 1,\mp \frac{c_{\mu }\mu ^{2}}{c_{2}m_{2}^{2}}\right) ^{T}, \\
\Phi _{0}^{(2\pm )} &:&=2\sqrt{\frac{c_{3}c_{\mu }m_{3}^{2}\mu
^{4}+c_{1}m_{1}^{2}\kappa }{g\kappa }}\left( \pm 1,\frac{c_{3}c_{\mu
}m_{3}^{2}\mu ^{2}}{\kappa },\mp 1,\frac{c_{\nu }c_{\mu }\nu ^{2}\mu ^{2}}{%
\kappa },0\right) ^{T}.
\end{eqnarray}%
Expanding around these vacua the corresponding squared mass matrices are 
\begin{equation}
M_{i}^{2}=\left( 
\begin{array}{ccccc}
A_{i} & -c_{\mu }\mu ^{2} & 0 & 0 & 0 \\ 
c_{\mu }\mu ^{2} & -c_{2}m_{2}^{2} & c_{\nu }\nu ^{2} & 0 & 0 \\ 
0 & -c_{\nu }\nu ^{2} & -c_{3}m_{3}^{2} & 0 & 0 \\ 
0 & 0 & 0 & B_{i} & c_{\mu }\mu ^{2} \\ 
0 & 0 & 0 & -c_{\mu }\mu ^{2} & -c_{2}m_{2}^{2}%
\end{array}%
\right) ,\qquad i=0,1,2,
\end{equation}%
with%
\begin{equation}
A_{0}=B_{0}=-c_{1}m_{1}^{2},~A_{1}=\frac{\mu ^{4}}{c_{2}m_{2}^{2}}%
,~B_{1}=2c_{1}m_{1}^{2}+3A_{1},~A_{2}=2c_{1}m_{1}^{2}+3B_{2},~B_{2}=\frac{%
c_{3}m_{3}^{2}\mu ^{4}}{\kappa }.
\end{equation}%
The different signs in $\Phi _{0}^{(1\pm )}$ and $\Phi _{0}^{(2\pm )}$ give
rise to the same mass matrix so that we may ignore that distinction in what
follows.

The parameter study of all mass matrices $M_{i}$ reveals that there are
physical regions for all three models bounded by exceptional points
similarly as in the previous section for the purely complex $\mathcal{I}_{3}$%
-model. Our crucial observation is here that the determinants%
\begin{eqnarray}
\det M_{0}^{2} &=&-(c_{1}c_{2}m_{1}^{2}m_{2}^{2}+\mu
^{4})(c_{1}m_{1}^{2}\kappa +c_{3}m_{3}^{2}\mu ^{4}), \\
\det M_{1}^{2} &=&-\frac{2\mu ^{4}\nu ^{4}}{c_{2}m_{2}^{2}}%
(c_{1}c_{2}m_{1}^{2}m_{2}^{2}+\mu ^{4}), \\
\det M_{2}^{2} &=&\frac{2\mu ^{4}\nu ^{4}}{\kappa }(c_{1}m_{1}^{2}\kappa
+c_{3}m_{3}^{2}\mu ^{4}),
\end{eqnarray}%
are always nonvanishing when $m_{i}\neq 0$, $\mu \neq 0$ and $\nu \neq 0$.
Hence in all sectors of the $\mathcal{PT}$-symmetries this model does not
possess any Goldstone boson, which is expected in the absence of a global
symmetry. There are of course special points as for the previous model, such
as $\mu _{s}^{4}=-c_{1}c_{2}m_{1}^{2}m_{2}^{2}$ or $\bar{\mu}%
_{s}^{4}=-c_{1}m_{1}^{2}\kappa /c_{3}m_{3}^{2}$, for which massless bosons
enter the model. However, these massless bosons are present in the model
from the very beginning and not the result of the breaking of a continuous
symmetry by expanding around particular vacua. Hence they are not
interpreted as Goldstone bosons.

\section{General interaction term}

In our initial Lagrangian density functional (\ref{actionan}) we chose a
particularly simple interaction term and carried out our analysis for an
even simpler version. In this section we explore the possibilities of
allowing for more general interaction terms so that the action still
respects the discrete $\mathcal{CPT}$-symmetries (\ref{CPT12}) and the
continuous global $U(1)$-symmetry (\ref{U11}), while keeping the kinetic and
mass term as previously. We present here explicitly the case for $\mathcal{I}%
_{3}$, after which it becomes evident how to generalize to all $\mathcal{I}%
_{n}$. We carry out our analysis for the equivalent action $\hat{{\mathcal{I}%
}}_{n}$.

We find that the action 
\begin{equation}
\hat{{\mathcal{I}}}_{3}\left[ \Phi \right] =\frac{1}{2}\int d^{4}x\left[
\partial _{\mu }\Phi ^{T}I\partial ^{\mu }\Phi -\Phi ^{T}H\Phi -\frac{g}{8}%
\left( \Phi ^{T}E\Phi \right) ^{2}-\frac{g}{8}\left( \Phi ^{T}F\Phi \right)
^{2}\right] ,  \label{gencpt}
\end{equation}%
is $\widehat{\mathcal{CPT}}$ and $U(1)$-invariant, where we recalled the
field vector $\Phi :=(\varphi _{1},\chi _{2},\varphi _{3},\chi _{1},\varphi
_{2},\chi _{3})^{T}$ and introduced 
\begin{equation}
H=\left( 
\begin{array}{cccccc}
-c_{1}m_{1}^{2} & c_{\mu }\mu ^{2} & 0 & 0 & 0 & 0 \\ 
c_{\mu }\mu ^{2} & c_{2}m_{2}^{2} & c_{\nu }\nu ^{2} & 0 & 0 & 0 \\ 
0 & c_{\nu }\nu ^{2} & -c_{3}m_{3}^{2} & 0 & 0 & 0 \\ 
0 & 0 & 0 & -c_{1}m_{1}^{2} & -c_{\mu }\mu ^{2} & 0 \\ 
0 & 0 & 0 & -c_{\mu }\mu ^{2} & c_{2}m_{2}^{2} & -c_{\nu }\nu ^{2} \\ 
0 & 0 & 0 & 0 & -c_{\nu }\nu ^{2} & -c_{3}m_{3}^{2}%
\end{array}%
\right) ,E=\left( 
\begin{array}{cc}
A & 0 \\ 
0 & \Omega A\Omega%
\end{array}%
\right) ,F=\left( 
\begin{array}{cc}
0 & B \\ 
\Omega B\Omega & 0%
\end{array}%
\right) .
\end{equation}%
Here $A$ and $B$ can be arbitrary $3\times 3$-matrices and $\limfunc{diag}%
\Omega =(-1,1,-1)$.

We briefly show how the form of this action is obtained. \ The respective
symmetries (\ref{cpthat}) and (\ref{U11}) are realised as follows 
\begin{eqnarray}
\widehat{\mathcal{CPT}}_{1,2} &:&~\hat{{\mathcal{I}}}_{3}\left[ \Phi \right]
=\hat{{\mathcal{I}}}_{3}\left[ C_{1,2}\Phi \right] \\
U(1) &:&~\hat{{\mathcal{I}}}_{3}\left[ \Phi \right] =\hat{{\mathcal{I}}}_{3}%
\left[ U\Phi \right]
\end{eqnarray}%
with%
\begin{equation}
C_{1,2}=\pm \left( 
\begin{array}{cc}
\mathbb{I}_{3} & 0 \\ 
0 & -\mathbb{I}_{3}%
\end{array}%
\right) ,~~~\ U=\mathbb{I}_{6}+\alpha \hat{\Omega}=\mathbb{I}_{6}+\alpha
\left( 
\begin{array}{cc}
0 & \Omega \\ 
-\Omega & 0%
\end{array}%
\right) ,  \label{CU}
\end{equation}%
when $\alpha $ is taken to be small. Next we compute how these symmetries
are implemented when taking the interaction term to be of the general form%
\begin{equation}
\frac{g}{16}\left( \Phi ^{T}\hat{E}\Phi \right) ^{2},~~~\ \hat{E}=\left( 
\begin{array}{cc}
A & B \\ 
C & D%
\end{array}%
\right) ,  \label{int}
\end{equation}%
with as yet unknown $3\times 3$-matrices $A$, $B$, $C$ and $D$. The
transformed Noether current (\ref{JN}) resulting from the $U(1)$-symmetry (%
\ref{CU}) 
\begin{equation}
j_{\mu }=\frac{\alpha }{2}\left( \partial _{\mu }\Phi ^{T}\hat{\Omega}\Phi
-\Phi ^{T}\hat{\Omega}\partial _{\mu }\Phi \right)
\end{equation}%
is vanishing upon using the equation of motion for the action $\hat{{%
\mathcal{I}}}_{3}\left[ \Phi \right] $ with interaction term (\ref{int}) 
\begin{equation}
-\square \Phi -H\Phi -\frac{g}{4}\left( \Phi ^{T}\hat{E}\Phi \right) \hat{E}%
\Phi =0
\end{equation}%
if 
\begin{equation}
\partial _{\mu }j^{\mu }=\frac{\alpha }{2}\left( \square \Phi ^{T}\hat{\Omega%
}\Phi -\Phi ^{T}\hat{\Omega}\square \Phi \right) =\frac{\alpha }{2}\Phi
^{T}\left( \left[ \hat{\Omega},H\right] -\frac{g}{4}\Phi ^{T}\hat{E}\Phi %
\left[ \hat{\Omega},\hat{E}\right] \right) \Phi =0.
\end{equation}%
Combining the constraints for the $\widehat{\mathcal{CPT}}$ and $U(1)$%
-symmetry we require therefore 
\begin{equation}
\left[ \hat{\Omega},H\right] =0,\quad \left[ \hat{\Omega},\hat{E}\right]
=0,\quad \left[ C_{1,2},H\right] =0,\quad \left[ C_{1,2},\hat{E}\right] =0,
\label{con1}
\end{equation}%
or%
\begin{equation}
\left[ \hat{\Omega},H\right] =0,\quad \left[ \hat{\Omega},\hat{E}\right]
=0,\quad \left[ C_{1,2},H\right] =0,\quad \left\{ C_{1,2},\hat{E}\right\} =0,
\label{con2}
\end{equation}%
with $\left\{ \cdot ,\cdot \right\} $ denoting the anti-commutator. The
solutions to (\ref{con1}) for $\widehat{\mathcal{CPT}}_{1}$ and $\widehat{%
\mathcal{CPT}}_{2}$ are $E$ and $F$, respectively, whereas the solutions to (%
\ref{con2}) for $\widehat{\mathcal{CPT}}_{1}$ and $\widehat{\mathcal{CPT}}%
_{2}$ are $F$ and $E$, respectively. This mean the action (\ref{gencpt})
contains the most general $\widehat{\mathcal{CPT}}_{1,2}$ and $U(1)$
invariant interaction terms of the form (\ref{int}). There is no distinction
between a $\widehat{\mathcal{CPT}}_{1}$ or $\widehat{\mathcal{CPT}}_{2}$%
-invariant action as the solutions of (\ref{con1}) and (\ref{con2}) always
combine to allow for both $\widehat{\mathcal{CPT}}$-symmetries to be
implemented.

We carried out our analysis for the Goldstone boson for $\limfunc{diag}%
A=(1,0,0)$ and $B=0$, but from the above it is now evident how this
structure of more complicated interaction terms generalises to $\hat{{%
\mathcal{I}}}_{n}$, and therefore $\mathcal{I}_{n}$, for $n>3$. Similar
computations can also be carried out for the symmetries $\mathcal{CPT}_{3/4}$
and $\mathcal{CP}^{\prime }\mathcal{T}$, where $\mathcal{P}^{\prime }$ is
any of the six remaining operators constructed in section 3.5. We note here
that while it is a uniquly well defined process to identify the $\widehat{%
\mathcal{CPT}}$-symmetries when given the $\mathcal{CPT}$-symmetries, that
is going from $\mathcal{I}_{n}$ to $\hat{{\mathcal{I}}}_{n}$, care needs to
be taken in the inverse procedure.

\section{Conclusions and outlook}

We proposed and analysed a new non-Hermitian model with $n$ complex scalar
fields that possess a global $U(1)$-symmetry when none of the scalar fields
involved are self-conjugate. Making use of the general fact that actions can
be similarity transformed without changing the content of the theory, as
long as the equal time-commutation relations are preserved, we mapped the
models to equivalent Hermitian systems. The models obtained in this manner
possess different types of vacua that may either respect or break the global
continuous symmetry. As expected from the Hermitian version of Goldstone's
theorem the models do not possess any Goldstone bosons when the vacuum
around which the theory is expanded preserves the $U(1)$ symmetry, see
section 3.3, and when the symmetry is broken already on the level of the
action by taking some of the complex fields to be real, see section 4. In
both cases there are special points in the parameter space for which the
model contains massless particles, which are, however, not identified as
Goldstone bosons. In contrast, when expanding the action around a $U(1)$%
-symmetry breaking vacuum a Goldstone boson emerges. In the $\mathcal{PT}$%
-symmetric regime and at the standard exceptional point its explicit form in
terms of the original fields in the model can be identified, although it
takes on different forms in these two regimes. In contrast, at the
zero-exceptional point one can not identify the Goldstone boson, but only a
linear combination of it with another massless particle. Hence the general
statement of the Goldstone theorem holds for Hermitian as well as for
non-Hermitian actions, but the latter possesses special regimes with
behaviour that have no analogue in the former. As the reality of the mass
spectra and the explicit form of the Goldstone bosons are strictly governed
by the $\mathcal{PT}$-symmetric at the tree approximation this leads to the
interesting possibility that one may have models with broken $\mathcal{CPT}$%
-symmetry on the level of the action, but with real physical masses.

There are various issues that are worthy further investigation. First of all
one may of course consider more complicated complex models by investigating
those for larger values of $n$ and also include more involved interaction
terms as derived in section 5. In particular, one may construct those that
remain $\mathcal{CPT}$-symmetric beyond the tree level when employing the
remaining six $\mathcal{P}$-operators constructed in section 3.5. A richer
structure is expected to be revealed by considering non-Hermitian models
that possess global continuous non-Abelian symmetries so that more Goldstone
bosons are generated via a symmetry breaking \cite{AFTTprep}.

\newif\ifabfull\abfulltrue

\end{document}